\documentclass[11pt,preprint2]{aastex}

\usepackage[T1]{fontenc}
\usepackage{lmodern}
\usepackage{breqn}
\usepackage{graphicx}
\usepackage{txfonts}
\usepackage{lscape}
\usepackage{aas_macros}
\usepackage{amssymb}
\usepackage{natbib}
\bibpunct{(}{)}{;}{a}{,}{,}
\usepackage{longtable}
\usepackage[hang,bf,footnotesize]{subfigure}
\usepackage[figuresright]{rotating}
\begin{document}

\title{Accretion and outflow in the proplyd-like objects near Cygnus OB2}

\author{M. G. Guarcello\altaffilmark{1,2}, J. J. Drake\altaffilmark{2}, N. J. Wright\altaffilmark{3,2}, D. Garc\'{i}a-Alvarez\altaffilmark{4,5,6}, K. E. Kraemer\altaffilmark{7}}

\altaffiltext{1}{INAF - Osservatorio Astronomico di Palermo, Piazza del Parlamento 1, I-90134, Palermo, Italy }
\altaffiltext{2}{Smithsonian Astrophysical Observatory, MS-67, 60 Garden Street, Cambridge, MA 02138, USA}
\altaffiltext{3}{CAR/STRI, University of Hertfordshire, College Lane, Hatfield, AL10 9AB, UK}
\altaffiltext{4}{Dpto. de Astrof\'{i}sica, Universidad de La Laguna, 38206 - La Laguna, Tenerife, Spain}
\altaffiltext{5}{Grantecan CALP, 38712 Bre\~{n}a Baja, La Palma, Spain}
\altaffiltext{6}{Instituto de Astrof\'{i}sica de Canarias, E-38205 La Laguna, Tenerife, Spain}
\altaffiltext{7}{Institute for Scientific Research, Boston College, Kenny Cottle L106B, Newton, MA 02459-1161, USA}

\begin{abstract}

Cygnus~OB2 is the most massive association within $2\,$kpc from the Sun, hosting hundreds of massive stars, thousands of young low mass members, and some sights of active star formation in the surrounding cloud. Recently, 10 photoevaporating proplyd-like objects with tadpole-shaped morphology were discovered in the outskirts of the OB association, approximately 6-14$\,$pc away from its center. The classification of these objects is ambiguous, being either evaporating residuals of the parental cloud which are hosting a protostar inside, or disk-bearing stars with an evaporating disk, such as the evaporating proplyds observed in the Trapezium Cluster in Orion. In this paper we present a study based on low resolution optical spectroscopic observations made with the Optical System for Imaging and low Resolution Integrated Spectroscopy (OSIRIS), mounted on the $10.4\,$m Gran Telescopio CANARIAS (GTC), of two of these protostars. The spectrum of one of the objects shows evidence of accretion but not of outflows. In the latter object, the spectra show several emission lines indicating the presence of an actively accreting disk with outflow. We present estimates of the mass loss rate and the accretion rate from the disk, showing that the former exceeds the latter as observed in other known objects with evaporating disks. We also show evidence of a strong variability in the integrated flux observed in these objects, as well as in the accretion and outflow diagnostics. 

\end{abstract}


\section{Introduction}
\label{intro}

Star formation in our Galaxy occurs in a variety of environments, with one of the main modes being in clusters hosting OB stars, and then characterized by an intense UV radiation field. In the past decade, several works have been focused on the understanding of how the evolution of circumstellar disks and the processes of star and planet formation are affected by an intense local UV field \citep{JohnstoneHB1998,AdamsHLG2004,ThroopBally2005,GuarcelloMPP2010}. \par

  In low mass environments, characterized by a small number of massive stars and weak local UV field, disks are observed to dissipate in less than $10\,$Myrs. In such an environment, the fraction of cluster members holding an inner disk declines exponentially with an e-folding time of $2.5\,$Myrs \citep{Mamajek2009}. \par

  In more massive environments, the hosted OB stars, despite their short lifetimes, have a strong impact on the evolution of the parental cloud, and on nearby young protostars and their circumstellar disks. The intense UV and X-ray radiation from OB stars ionizes the surrounding cloud turning it into a HII region and clearing a cavity delimited by an expanding ionization front \citep{Whitworth1979}. When the expanding front sweeps over local overdensities in the cloud, a residual evaporating globule of gas (called ``Evaporating Gaseous Globules'', EGGs) can be produced \citep{HesterSSL1996}. The possibility that EGGs can actually be collapsing and forming one or more protostars in their center has been debated. For instance, in the Eagle Nebula, hosting tens of OB stars, only 15\% of the 73 EGGs observed close to the {\it pillars of creation} have protostars in their center \citep{McCaughreanAndersen2002}. \par

In general, new stars form in the contracting molecular cloud during the expansion of the ionization front created by the massive stars \citep{Elmegreen2011}. As soon as these young stars emerge from the expanding front, they are exposed to the ionizing radiation emitted by the massive stars. As directly observed in the Trapezium in Orion \citep{HenneyODell1999} using the Hubble Space Telescope (HST), where an intense UV field is created by the O6V star $\Theta^1$ Ori, disks irradiated by massive stars can be quickly dissipated by photoevaporation. This process is caused by photons in the Extreme Ultraviolet (EUV) regime, with energy $h\nu\,>13.6\,$eV, and even more energetic X-ray photons that ionize the gas in the circumstellar disks, raising the disk temperature to more than $10^4\,$K; and photons in the Far Ultraviolet (FUV) regime, with energy between $6\,$eV$\,<\,h\nu\,<13.6\,$eV, that dissociate the gas molecules, heating the gas in the disk up to few $10^3\,$K. At these temperatures, the high thermal pressure in the disk drives a flow of gas away from its surface, depleting the disk of both gas and small dust grains \citep{JohnstoneHB1998,BalogRSM2006}. \par

  In the Trapezium in Orion, the externally induced photoevaporation is mainly driven by the FUV radiation, given that the EUV radiation is easily absorbed by the intervening material and the photoevaporative gas itself. In the latter case, the incident EUV radiation ionizes the outflowing gas at a distance from the disk surface, called the ``{\it ionization radius}'' ($r_{if}$) which depends on several variables, such as the intensity of the incident UV flux. This ionized outflowing gas forms an envelope with a characteristic cometary shape with a head pointed toward the ionizing source and molded by the incident radiation and stellar wind \citep{JohnstoneHB1998,StorzerHollenbach1999}. The evaporating protoplanetary disks ($proplyds$) observed in the Trapezium by the HST are at distances $\leq 0.5\,$pc from $\Theta^1$ Ori \citep{HenneyODell1999}. However, consequences of externally induced photoevaporation have been observed at larger distances in other clusters hosting more massive stars, such as in NGC~2244 \citep{BalogMRS2007}, NGC~6611 \citep{GuarcelloPMD2007,GuarcelloMDP2009,GuarcelloMPP2010}, Pismis~24 \citep{FangBKH2012}, and Cygnus~OB2 \citep{Guarcelloinprep1}. \par

Candidate evaporating proplyds also have been observed in other clusters: in NGC 2244, IC 1396, and NGC 2264 \citep{BalogRSM2006}, in W5 \citep{KoenigAKS2008}, in the Trifid Nebula \citep{Yusef-ZadehBG2005}, the Lagoon Nebula \citep{StecklumHFH1998}, the Carina star-forming region \citep{SmithBM2003}, NGC 3603 \citep{BrandnerGCD2000}, and Cygnus~OB2 \citep{WrightDDG2012, WrightPGD2014}. The latter two groups are the best candidates for being actual evaporating proplyds rather than EGGs. In the young and massive association Cygnus~OB2 \citep{WrightDDV2010} the 10 proplyd-like objects discovered by \citet{WrightDDG2012} analyzing H$\alpha$ images of this area obtained from the Isaac Newton Telescope (INT) Photometric H-Alpha Survey (IPHAS, \citealp{DrewGIA2005}). These objects are found at distances between $6\,$pc and $14\,$pc from the main group of O stars of the association. The size of their envelopes range from $18\times10^3$ to $113\times10^3\,$AU, which is much larger than those observed in Orion ($40-400\,$AU; \citealp{HenneyODell1999}). \citet{WrightDDG2012} suggested a classification of these objects as photoevaporating protostars, mainly given the presence of a central protostar in 70\% of the cases, some of which are classified as embedded stars with disks by \citet{GuarcelloDWD2013}.\par

  In this paper we study optical low-resolution spectra of the central protostar in the proplyd-like objects \#5 and \#7, using the nomenclature of \citet{WrightDDG2012}, to probe their nature and physical conditions and confirm their classification as stars with evaporating disks. The paper is organized as follows: The properties of the two objects are described in Sect. \ref{targets_sec}; data reduction is described in Sect. \ref{data_sec}. In Sect. \ref{central_sec} we classify the two central stars and study their circumstellar environments. The accretion, outflow and variability are studied and discussed in Sect. \ref{acou_sec}, \ref{varia_sec} and \ref{discussion}. 

\section{The targets}
\label{targets_sec}

\begin{figure*}[]
        \centering
        \includegraphics[width=7cm]{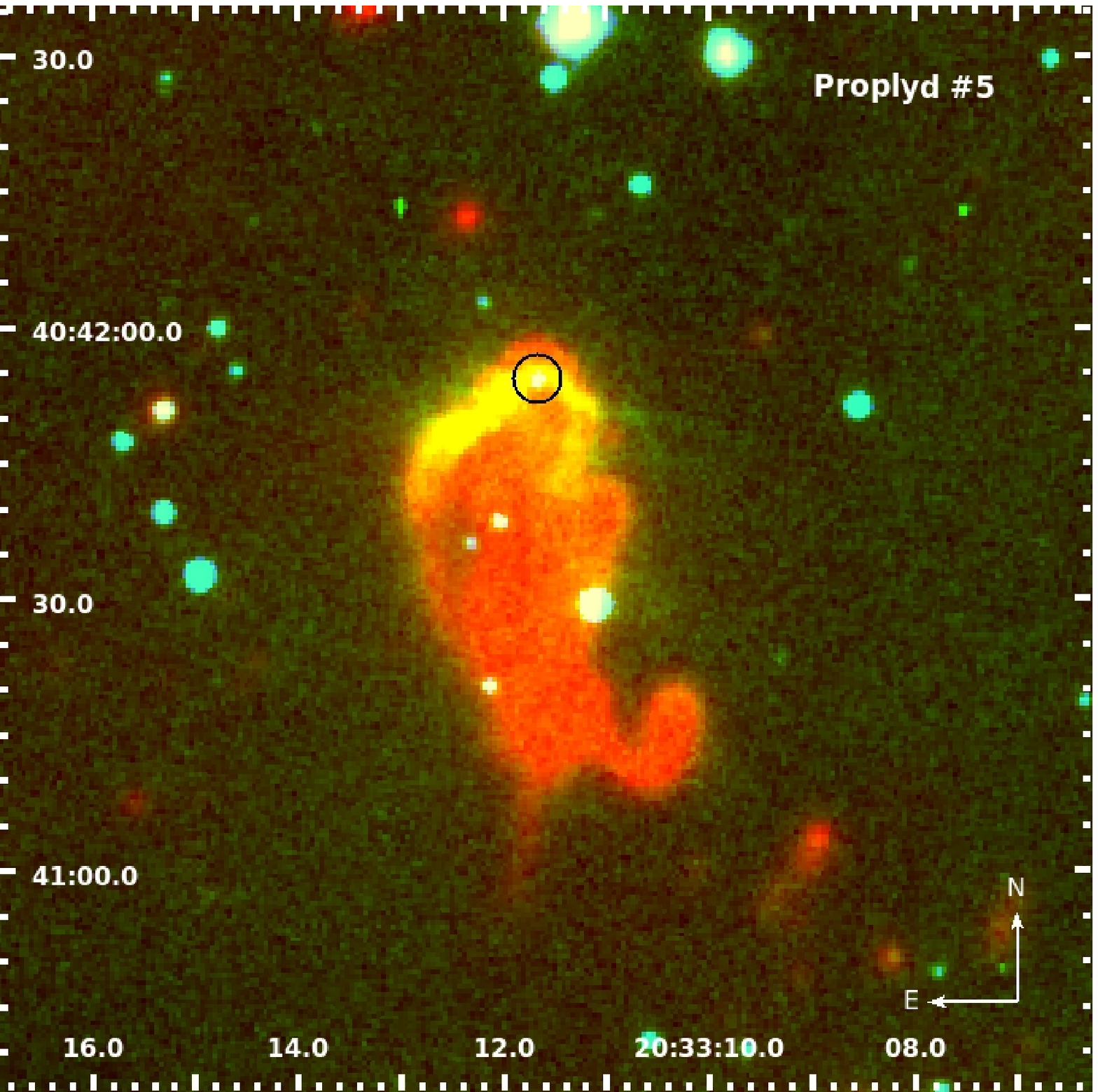}
        \par
        \includegraphics[width=7cm]{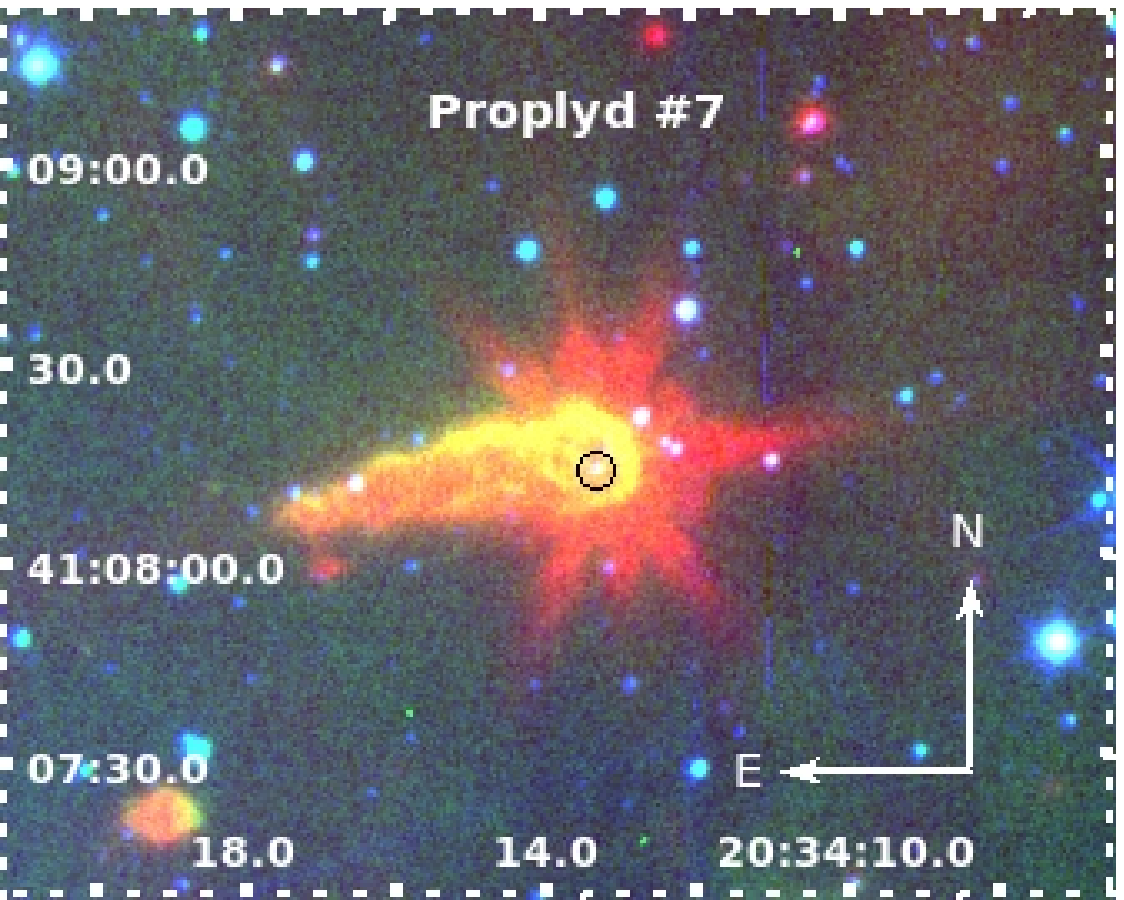}
        \caption{RGB images of the area surrounding protostar \#5 (upper panel) and \#7 (bottom panel). In both images, the emission at $8.0\,\mu$m from Spitzer is in red, in $H\alpha$ (from IPHAS) in green, in $r$ band in blue (from IPHAS in the upper panel and from OSIRIS in the bottom panel). The circles mark the position of the candidate embedded star observed with OSIRIS and IPHAS.}
        \label{field_img}
        \end{figure*}

Fig. \ref{field_img} shows two composite images in $H\alpha$ (green), $8.0\,\mu$m (red), and $r$ band (blue) of the two targets of this study. The H$\alpha$ data are taken from IPHAS, those at $8.0\,\mu$m from the {\it Spitzer Legacy Survey of the Cygnus X region} \citep{BeererKHG2010}, those in the $r$ band from IPHAS (upper panel) and observations from the Optical System for Imaging and low-Resolution Integrated Spectroscopy (OSIRIS) mounted on the $10\,m$ Gran Telescopio Canaria (GTC, \citealp{CepaAEG2000}),  shown in the bottom panel (see \citealp{GuarcelloWDG2012} for the analysis of the OSIRIS data of Cygnus~OB2). In both images, the position of the central star is marked with a circle. In this paper we assume a distance from the Sun of both objects of $1400\,$pc \citep{RyglBSM2012}. \par

\subsection{The protostar \#5}
\label{nmbr5_sec}

The envelope surrounding the protostar \#5 ($\alpha=20:33:11.66,\, \delta=40:41:54.3$) has a length of $75\times10^3\,$AU and a projected distance of $13.9\,$pc from the central cluster of O stars \citep{WrightDDG2012}. The central star has been identified by \citet{ComeronPRS2002} as a candidate OB star based on NIR photometry and low-resolution spectroscopy.  This object lies outside the field of the Chandra Cygnus~OB2 Legacy Project \citep{Drakeinprep} and thus it is not included in the list of stars with disks in Cygnus~OB2 compiled by \citet{GuarcelloDWD2013}. Its photometry (from the SDSS Data Release 9, IPHAS, UKIDSS public catalog, and IRAC from the Spitzer Legacy Survey of the Cygnus X region) is shown in Table \ref{prop5_tb}. It is a bright infrared source with colors in Spitzer bands typical of an embedded star with a circumstellar disk \citep{GutermuthMMA2008,GuarcelloDWD2013}.  \par

  The position of the embedded Young Stellar Object (YSO) corresponds to the ionization front in the envelope, although this may be a projection effect. This object is expected to be irradiated by a FUV flux of $1436\,$G$_0$\footnote{In terms of the Habing flux ``G$_0$'', equal to $1.6\times10^{-3}\,$erg$\,$cm$^{-2}\,$s$^{-1}$; $G_0=1.7$ corresponds to the average UV flux in the $912-2000\,$\AA{} spectral range in the Solar neighborhood \citep{Habing1968}} and an EUV flux of $3.9\times10^{10}\,$photons/s. These fluxes are calculated by projecting and summing the expected FUV and EUV fluxes emitted by each O star in Cyg~OB2 at the position of object \#5, using the intrinsic UV fluxes for O stars tabulated in \citet{ParravanoHM2003} and \citet{MartinSH2005} and their known spectral types \citep{Wrightinprep}. These fluxes are upper limits considering that we used the projected distances from the O stars and we ignored the intracluster absorption. For comparison, similar fluxes are experienced at distances of $1.3-1.4\,$pc from an O6V star such as $\Theta^1\,$Ori.\par

The protostar \#5 has been identified by \citet{GriffithHCB1991} as a $5\,$GHz continuum source. It has been named ``the Cradle Nebula'' by \citet{StrelnitskiBHS2013}, who identified it as a strong compact source in $^{12}$CO maps, and as a bright object in Spitzer and Herschel images, with different morphology at different wavelengths. In particular, the envelope presents a secondary clump on the western side that does not appear to be hosting a T~Tauri star. \par

\subsection{The protostar \#7}
\label{nmbr7_sec}

The protostar \#7 is located at $\alpha=20:34:13.27,\,\delta=41:08:13.8$, at a projected distance to the center of the association of $5.6\,$pc and with an envelope with an elongated structure $77\times10^3\,$AU long \citep{WrightDDG2012}. This object, also known as IRAS 20324+4057, was observed by the Hubble Space Telescope (PI: Sahai) with the Advanced Camera for Surveys using the broadband filters F606W ($\lambda_c = 5907\,$\AA{}) and F814W ($\lambda_c = 8333\,$\AA{}). A composite image of these two bands is shown in \citet{WrightDDG2012}. \par
 
  A study by \citet{PereiraMiranda2007} concluded that the envelope is a photoionized region rather than a shock-excited nebulosity. The central star has been classified by \citet{ComeronPRS2002} as a young massive star. In the bottom panel in Fig. \ref{field_img}, the central star is clearly detected using OSIRIS by \citet{GuarcelloDWD2013} in their search for the members of Cyg~OB2 with disks. The central star is number 1724 in their list of stars with disks, and it has been classified as a class~I object. Its optical counterpart is faint ($r=22.65^m \pm 0.03^m$), while the infrared counterpart is bright and red ($[3.6]=5.330^m\pm0.002^m$, $[3.6]-[4.5]=1.019^m\pm0.004^m$). \par

The envelope surrounding protostar \#7 has a canonical teardrop shape, with the embedded YSO at a distance of $\sim5^{\prime\prime}$ from the ionization front (about $7000\,$AU). This object is estimated to be irradiated by a FUV flux of $2802\,$G$_0$ and an EUV flux of $8.8\times10^{10}\,$photons/cm$^{2}$/s, enough to induce the photoevaporation of the circumstellar disk \citep{Guarcelloinprep1}\footnote{As shown by \citet{JohnstoneHB1998}, the disks of the proplyds in Orion are mainly irradiated by the FUV radiation. The EUV photons are in fact absorbed by the photoevaporating gas and the surrounding envelope, while copious FUV photons reach the disk. In this object, a bright ionization front is evident in the part of the envelope facing the O stars, so in principle the same process may be in action.}. An O6V star like $\Theta^1$Ori produces a similar flux at a distance of $0.9\,$pc, similar to the distance from $\Theta^1\,$Ori at which the evaporating proplyds in the Trapezium in Orion are observed.  \par

\section{Observations and data reduction}
\label{data_sec}

The two targets were observed with OSIRIS@GTC with a granted observing time of 4 hours (proposal GTC 28-12B). The log of the observations is shown in Table \ref{log_tb}. Observations were obtained on two different nights (2012-08-10$^{th}$, OB2, and -14$^{th}$, OB4), both with good seeing conditions. The protostar \#5 was observed using three different grisms: R1000B (3 exposures of $180\,$sec each), R2500R and R2500I (one $870\,$sec exposure with each grism); the protostar \#7 was observed 6 times using the R1000R grism, with exposures of $870\,$sec. We used slits with a width of $1.2^{\prime\prime}$ and the spectra were taken with $2\times 2$ binned pixels, as in the standard operation mode. Table \ref{grism_tb} lists the central wavelengths and spectral ranges of the grism we used. On each night of observation, arc lamp spectra, flat-field images, biases and spectra of spectrophotometric standards were taken. \par

  We reduced the spectra using standard IRAF routines\footnote{IRAF is distributed by the National Optical Astronomy Observatory, which is operated by the Association of Universities for Research in Astronomy (AURA) under cooperative agreement with the National Science Foundation.}. Spectra and calibration images were corrected for overscan, bias, and flat field using \emph{CCDPROC}. The bias images were combined using average and {\it minmax} rejection. We produced a master flat for each night by combining all the flat images available with the task \emph{RESPONSE}. The final wavelength and flux calibration was done using \emph{DOSLIT}. Given the highly variable diffuse emission around each target, no background subtraction was attempted. \par

\begin{figure*}[]
        \centering
        \includegraphics[width=6cm]{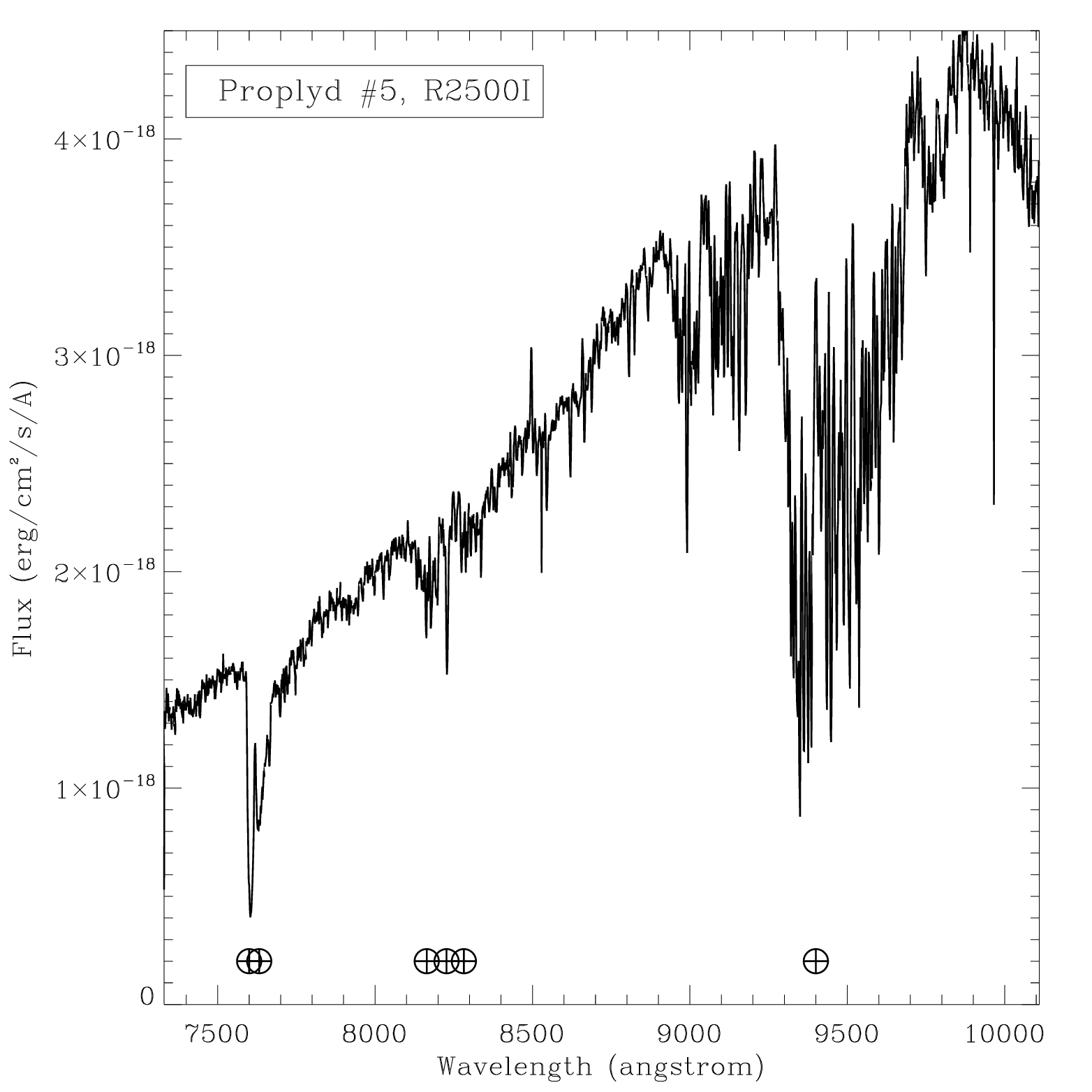}
        \includegraphics[width=6cm]{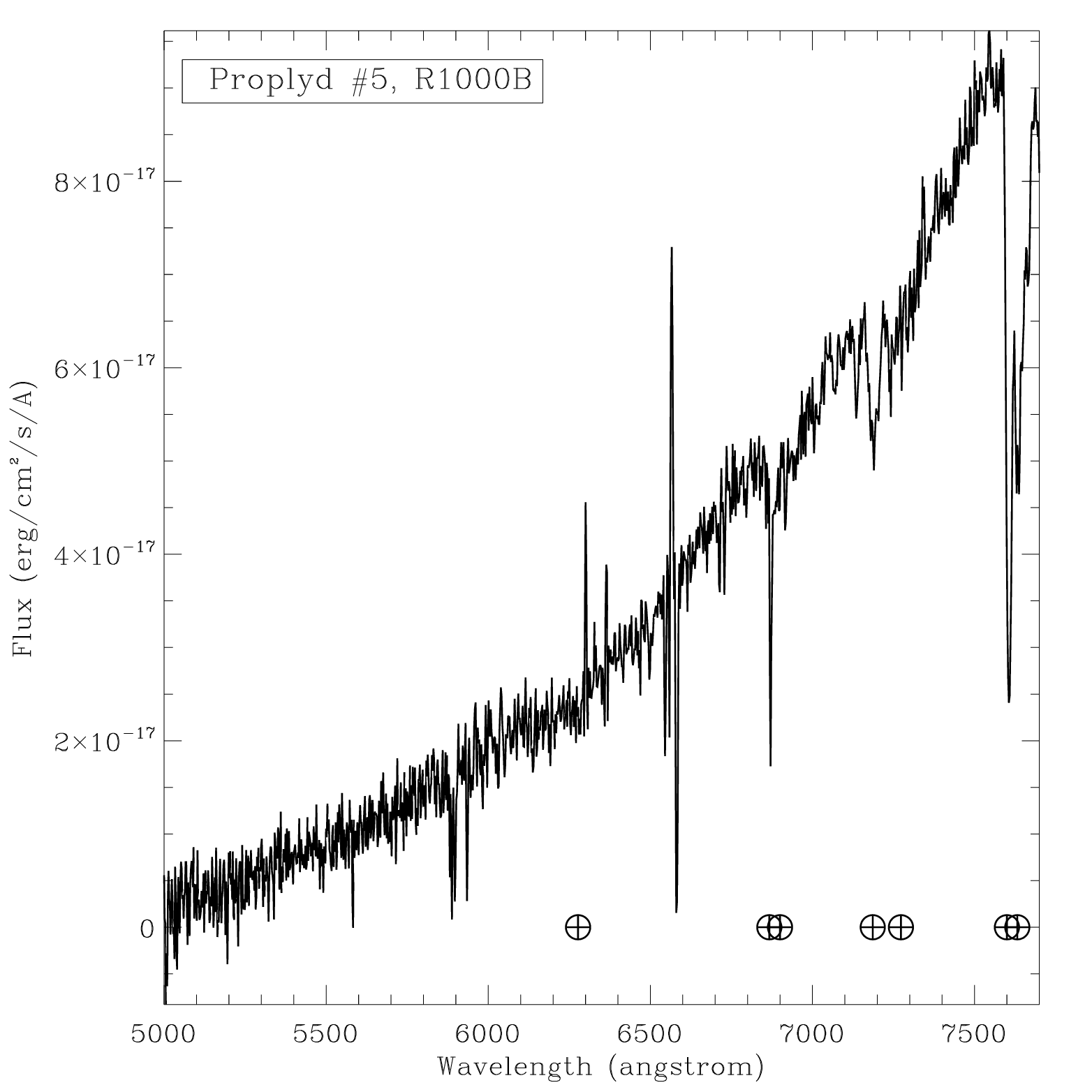}
        \par
        \includegraphics[width=6cm]{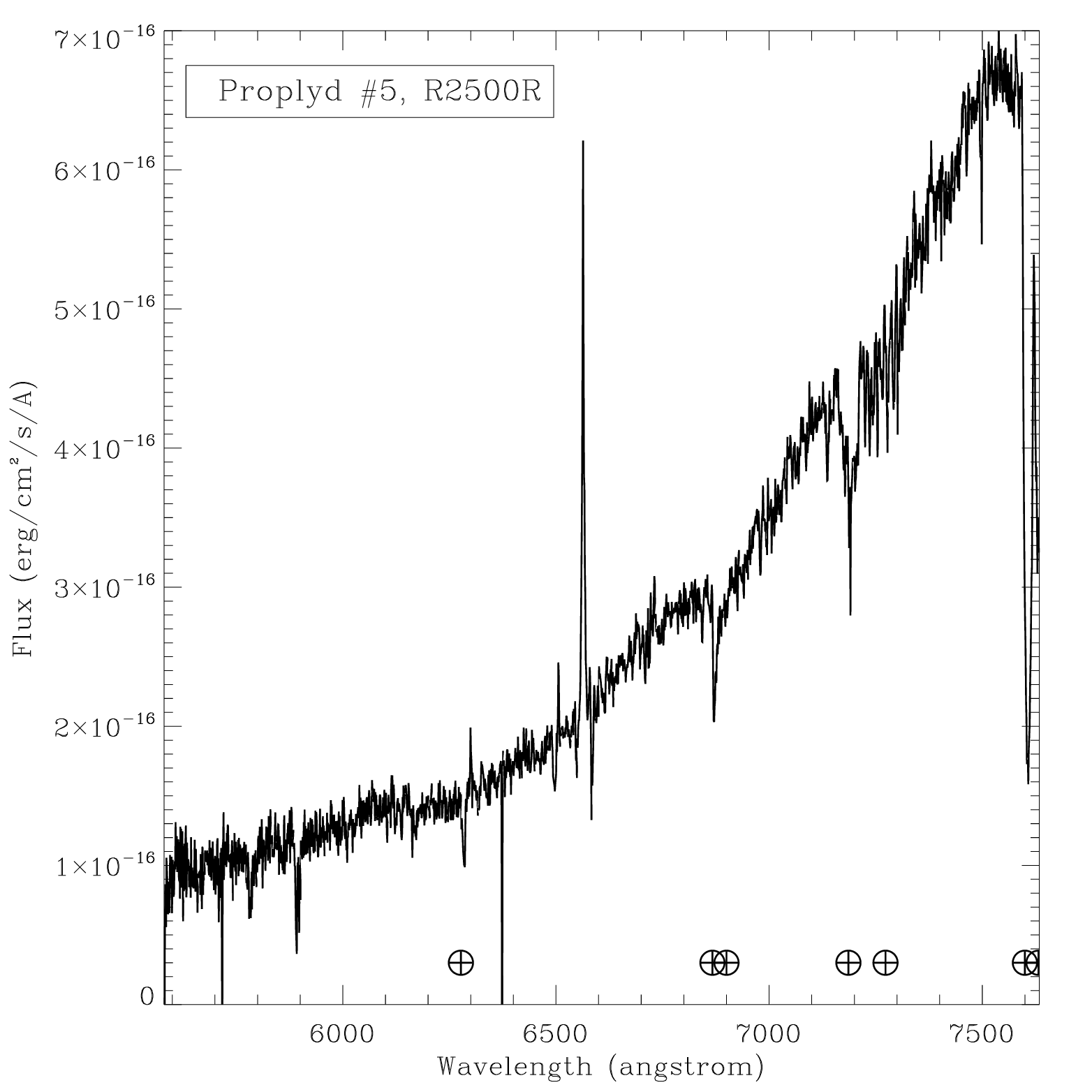}
        \includegraphics[width=6cm]{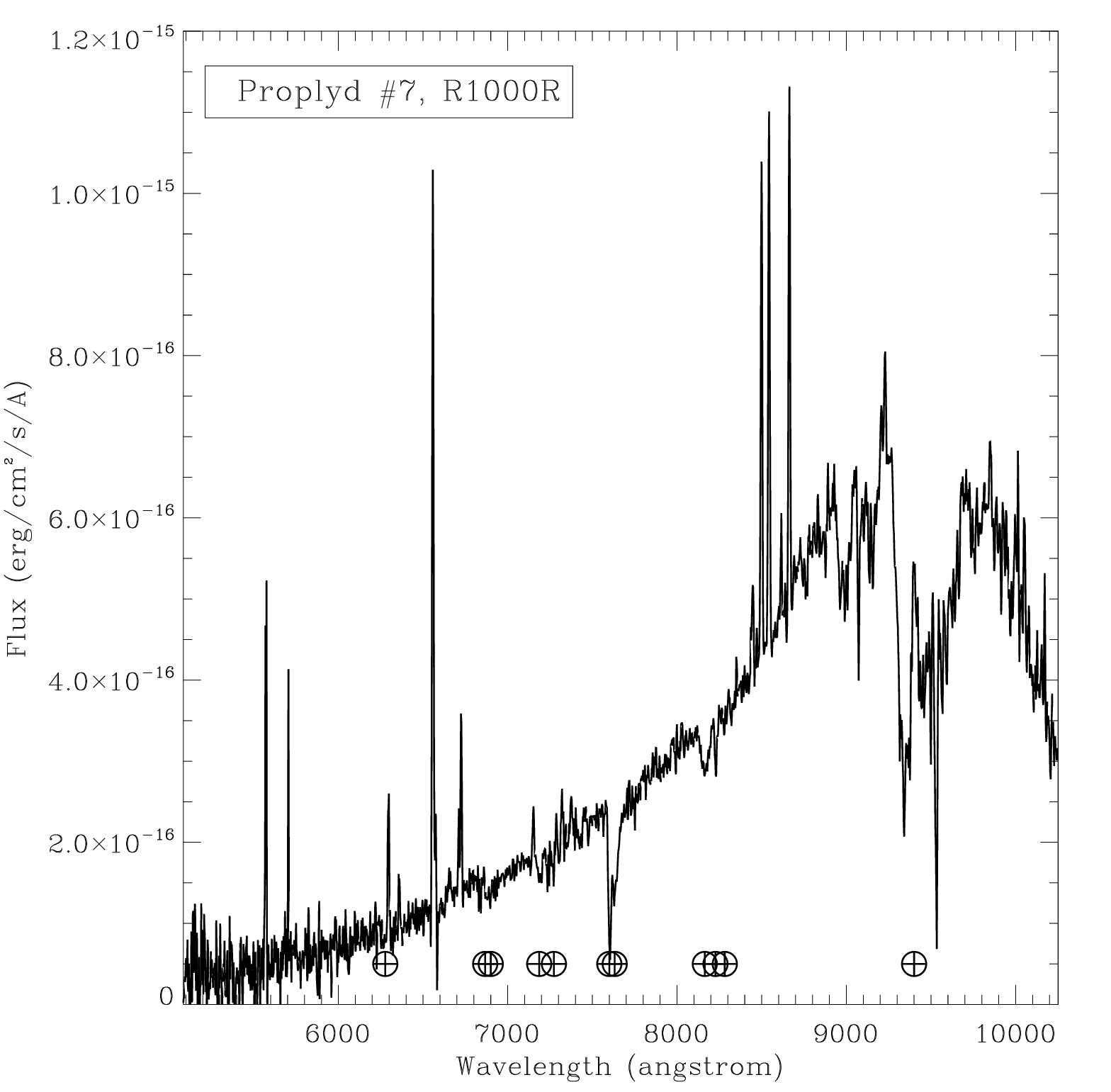}
        \caption{Reduced and calibrated, {\bf but not corrected for the extinction}, spectra of protostar \#5 (upper and lower left panels) and \#7 (lower right panel). The grism used and the telluric absorption features are also indicated.}
        \label{spectra_img}
        \end{figure*}

The GTC/OSIRIS spectra analyzed in this work are shown in Fig. \ref{spectra_img}, in which multiple spectra taken with the same grism are combined. The blue part ($\lambda<5000\,$\AA{}) of the R1000B spectrum of protostar \#5 is not shown here since it is completely dominated by noise. The atmospheric absorption features that in some cases dominate the spectra, such as at $\sim7600\,$\AA{} and $\sim9500\,$\AA{}, are also marked. Several emission lines, mainly from the protostar \#7, are evident. They provide excellent diagnostics for the infall/outflow process and the circumstellar environment and they will be analyzed in the next sections. Equivalent Widths (EW) and line fluxes were measured using the IRAF task \emph{SPLOT}. The spectra of both objects have been corrected for extinction using the IRAF task \emph{DEREDDEN} before measuring the line equivalent width and flux. We extracted a region of $\sim60\,$\AA{} around the line, and then normalized the extracted dereddened spectrum to the continuum using the task \emph{CONTINUUM}. EW and line fluxes were derived by fitting the line with a Voigt profile if $EW \ge 100\,$m\AA{} or with a Gaussian if $EW<100\,$m\AA{}.

\section{Properties of the central protostar}
\label{central_sec}

\subsection{Spectral classification}
\label{spec_sec}

As discussed in Sect. \ref{targets_sec}, the two protostars were classified as embedded OB stars by \citet{ComeronPRS2002} after analyzing their infrared spectra. We attempted a spectral classification for these two stars using the atlases compiled by \citet{WalbornFitzpatrick1990} and \citet{Torres-DodgenWeaver1993}, the classification of spectral standards by \citet{AllenStrom1995} and the \emph{SPTclass} code which uses the spectral classification scheme optimized for Herbig Ae/Be stars of \citet{HernandezCBH2004}. \par

  The spectrum of protostar \#7 does not show TiO bands or the Paschen series typical of KM and OBAF stars, respectively. For this reason, a spectral classification as a G star is more likely. However, the spectra of protostar \#7 do not show any particular absorption features typical of G stars, such as the NaI~$\lambda$5893 absorption line. This classification is thus still preliminary. We also compared the optical-infrared Spectral Energy Distribution (SED) of protostar \#7 with the grid of YSOs SED provided by \citet{RobitailleWI2007}. The observed SED fits models of both early and late stars, affected by an interstellar extinction ranging from $A_V\sim 9^m$ to $A_V\sim 12^m$. All the fitting models agree with an age younger than $1\,$Myr. Using the $1\,$Myr isochrone from 
  \citet{SiessDF2000}, and adopting the distance of $1400\,$pc, an extinction of about $A_V=12^m$ can be compatible with a G star, while smaller extinctions would be more compatible with later stars that we discard for the lack of observed TiO bands in the spectra. In the following, we adopt a G spectral type and we use an extinction of $A_V=12^m$ to deredden the spectra. However, we will show that our results are not qualitatively affected by adopting different values of extinction and stellar parameters.
  
The lack of absorption lines can be a consequence of veiling due to intense accretion from the disk. The accretion flow originates from the inner boundary of the disk, where the magnetic field funnels the gas that accretes onto the stellar surface \citep{UchidaShibata1984}. The accreting material falls at velocities of few hundred km/s, forming accretion spots at the footpoints of the open magnetic funnels. The energy released by the accretion process heats the accretion spots up to few $10^4\,K$. In this way, both the accreting material and the plasma in the accretion spots become sources of optical, UV, and soft X-ray radiation. This optical continuum emission, added to the photospheric emission, fills the photospheric absorption lines that consequently become weaker (e.g. \citealp{CalvetGullbring1998}). We will demonstrate that protostar \#7 is actively accreting, and the consequent veiling of its optical spectrum is a natural explanation for the lack of absorption lines. \par 

\begin{figure*}[]
        \centering
        \includegraphics[width=6cm]{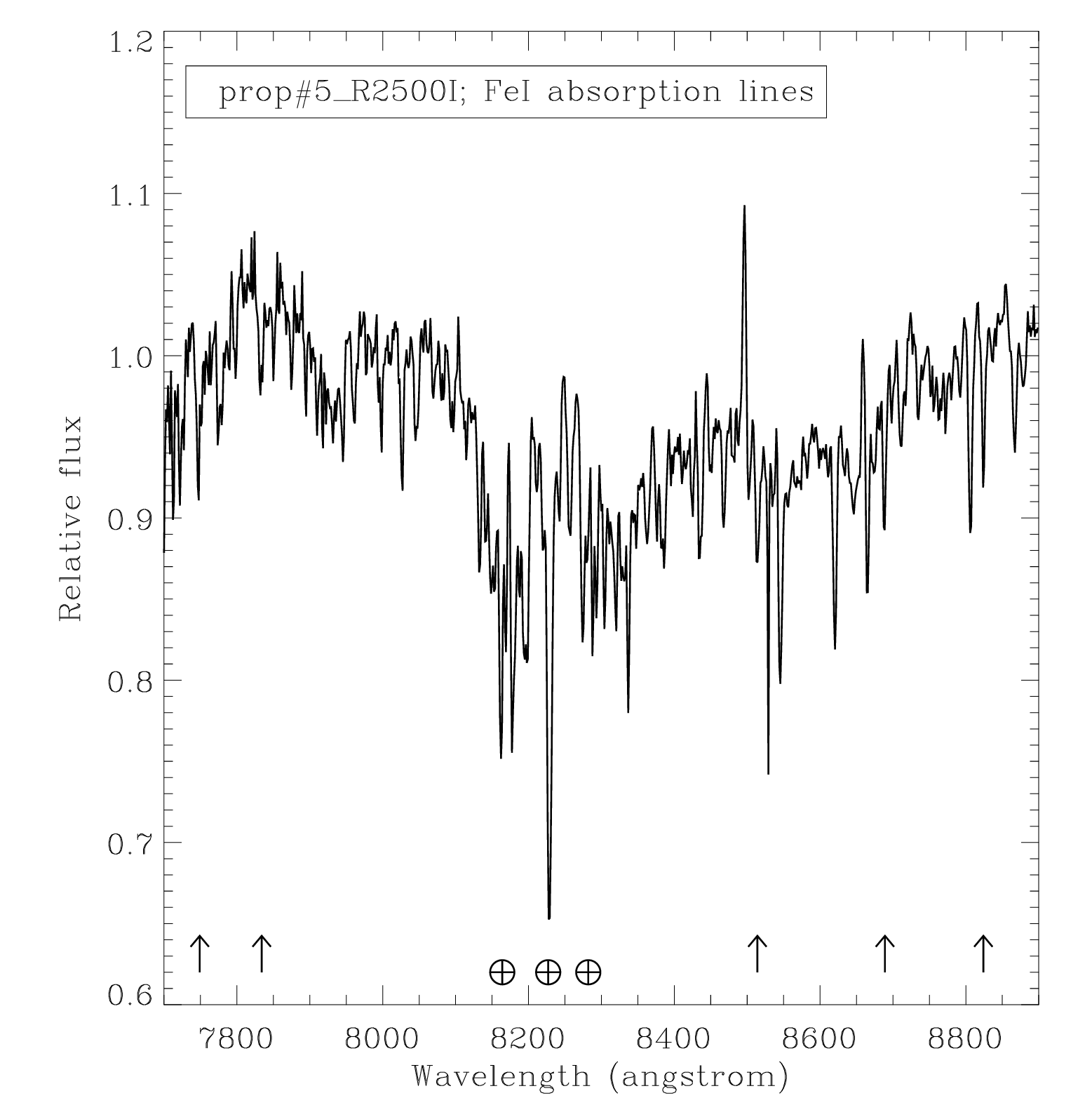}
        \includegraphics[width=6cm]{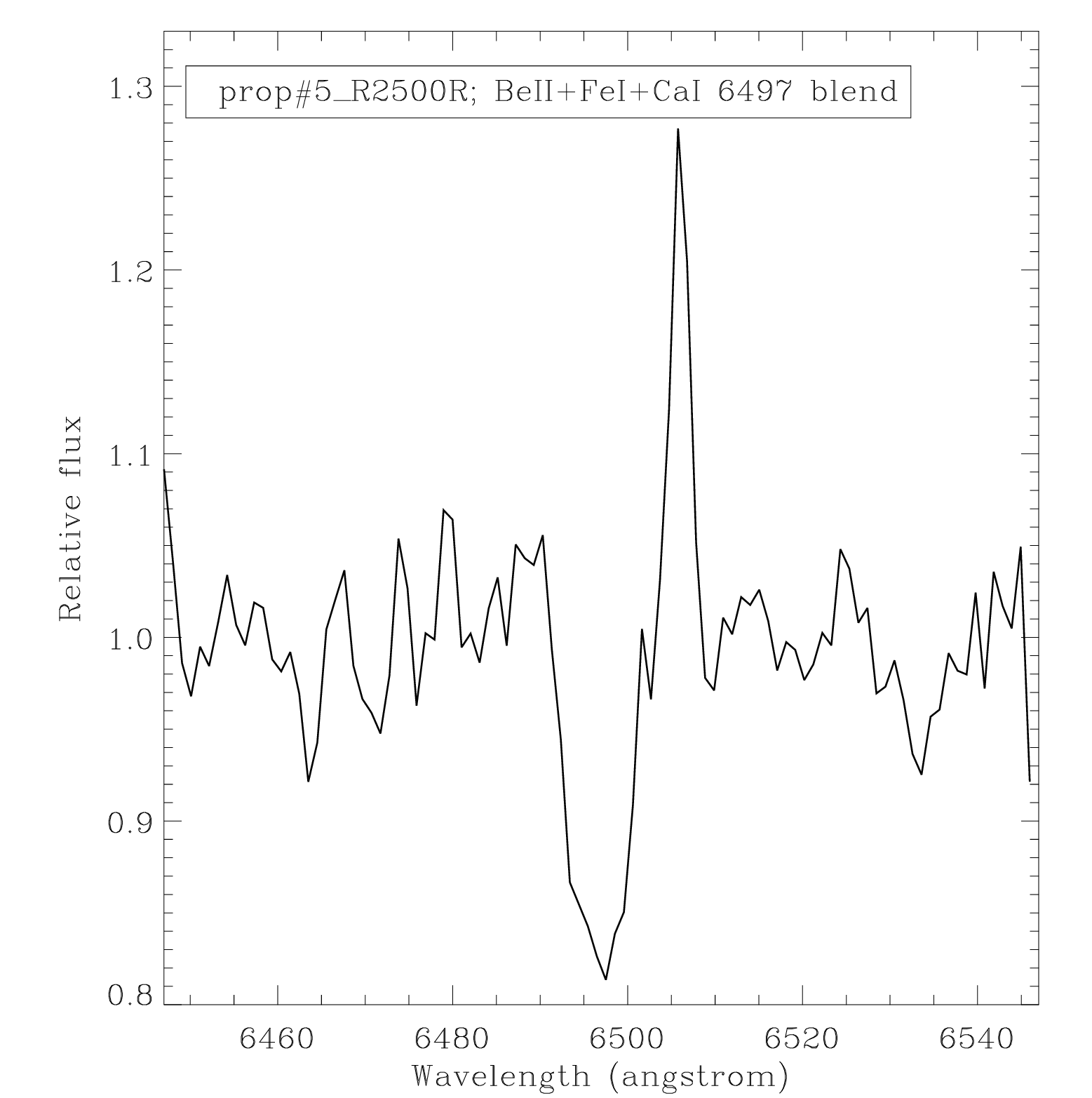}
        \par
        \includegraphics[width=6cm]{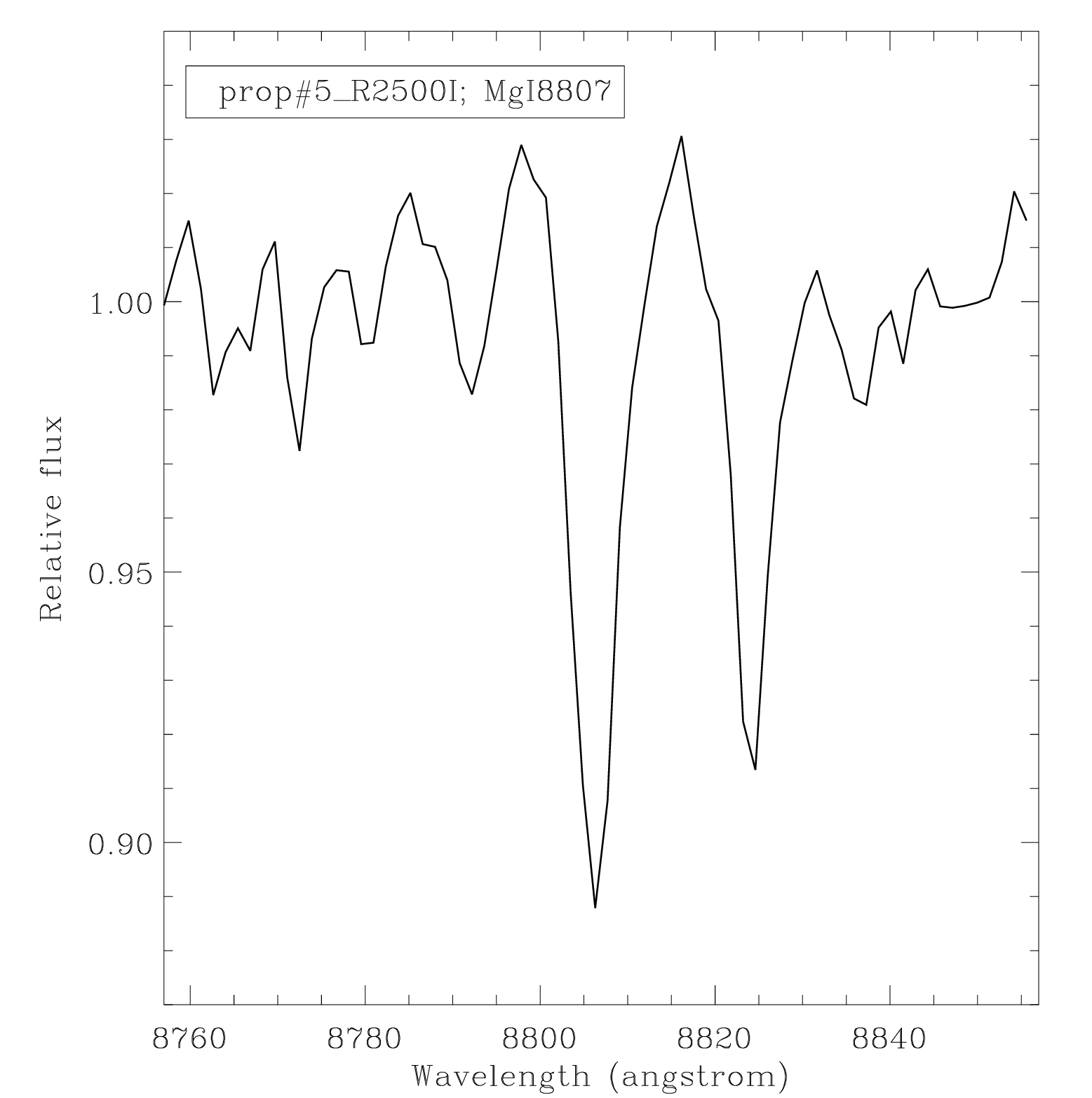}
        \includegraphics[width=6cm]{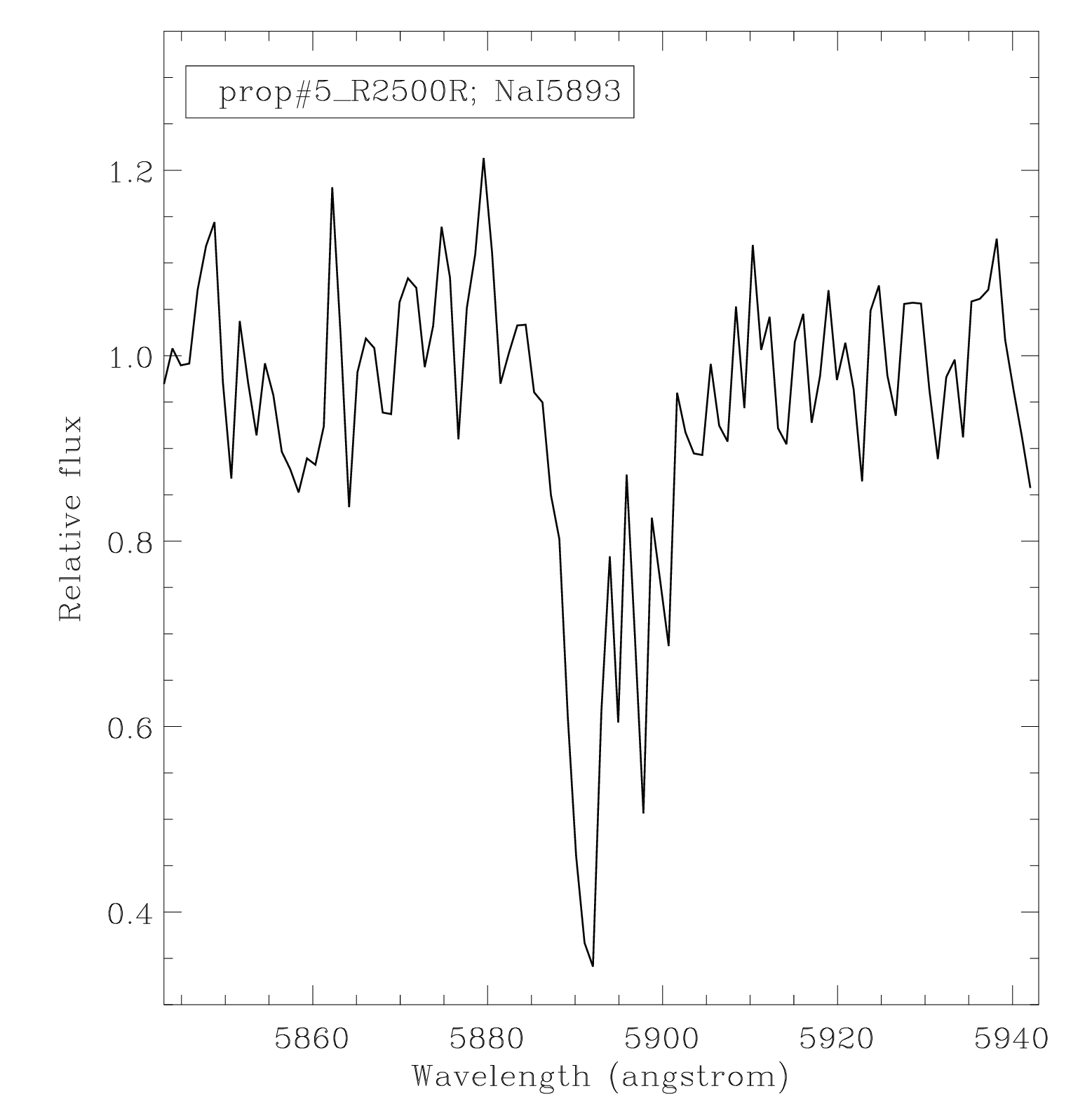}
        \caption{Lines used for the spectral classification of the protostar \#5: The complex of Fe~I absorption lines (marked with upward arrows), the Be~II+Fe~I+Ca~I~$\lambda$6497 blend, the Mg~I~$\lambda$8807 line, and the NaI~$\lambda$5893 line. All the spectra are normalized on the continuum.}
        \label{sptcl_img}
        \end{figure*}

  Several absorption lines are observed instead in the spectra of protostar \#5 (see Fig. \ref{sptcl_img}). The lack of Paschen lines in absorption excludes a spectral type earlier than G, while the lack of TiO bands excludes a spectral class later than late-K \citep{Torres-DodgenWeaver1993}. A G-early K spectral type is then the most likely solution. This is also supported by the presence of other absorption features such as several Fe~I lines between $7700\,$\AA{}  and $8900\,$\AA{}, the Be~II+Fe~I+Ca~I blend at $6497\,$\AA{}, and the MgI~$\lambda$8807 line. Furthermore, the EW of NaI~$\lambda$5893 is $5.8\pm0.5\,$\AA{}, which is typical of K2-K4 stars as indicated by SPTclass, and the EW of FeI~$\lambda$6495 is $1.45\pm0.05\,$\AA{}, indicating a spectral class earlier than K5. Merging these results, we adopt a spectral class of K$2\pm1$ for the protostar \#5, which corresponds to a mass larger than $2.2\,M_{\odot}$ for stars younger than $1\,Myr$ \citep{SiessDF2000}. Fig. \ref{sptcl_img} shows the lines that we used for the spectral classification of this star. \par

\subsection{Evidence of youth}
\label{lit_sec}

\begin{figure}[]
        \centering
        \includegraphics[width=6cm]{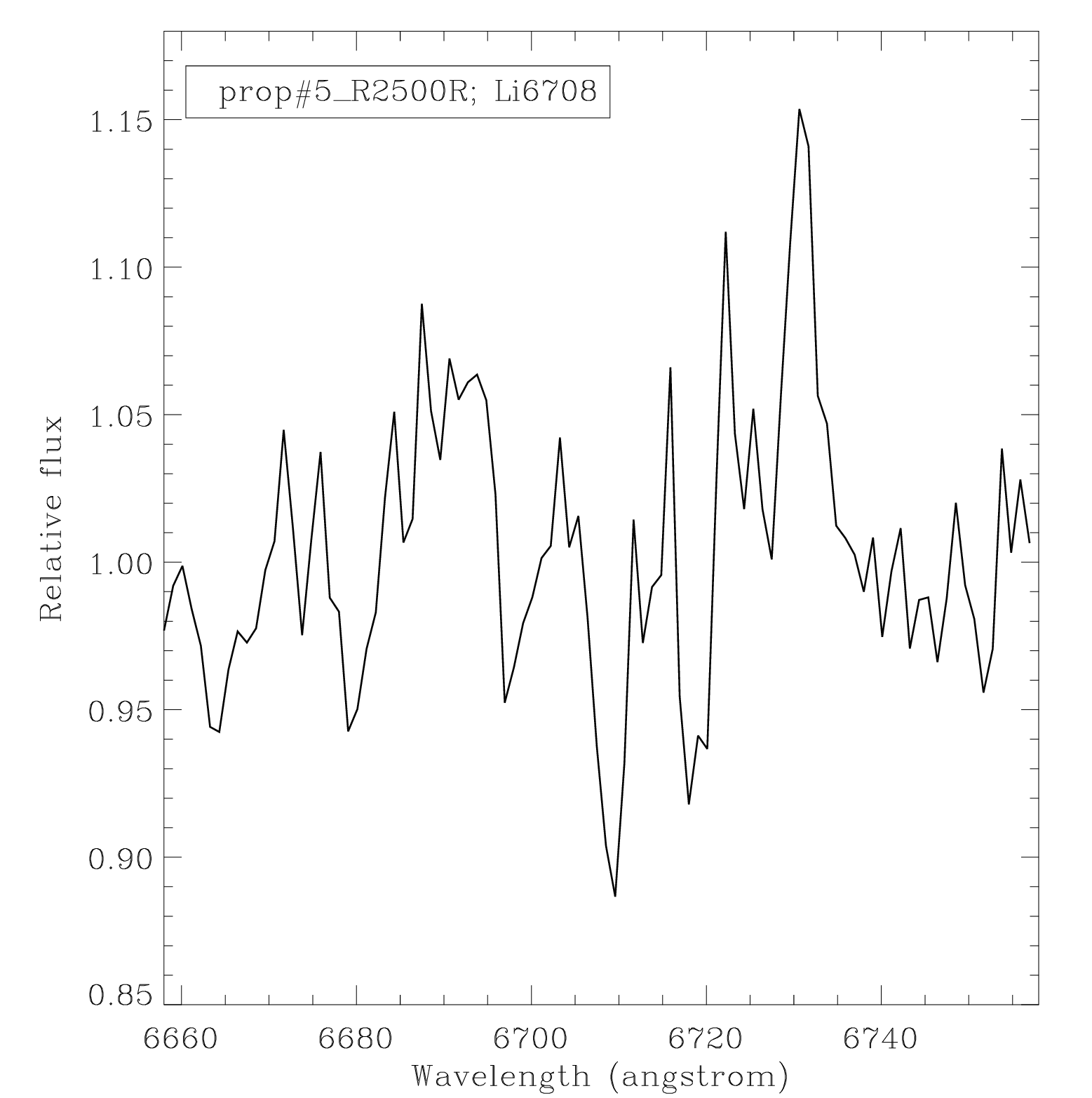}
        \caption{R2500R spectrum of the protostar \#5 around the faint absorption feature that may correspond to the Li~$\lambda$6708 absorption line (in the center).}
        \label{lit_img}
        \end{figure}

 A Li~$\lambda$6708 absorption line with an $EW\sim0.46\,$\AA{} is present in the spectra of the protostar \#5 (see Fig. \ref{lit_img}). This line is usually adopted as a criterion for selecting young stars. Lithium is burned at relatively low temperatures ($2.5 - 3.0\times 10^6\,K$), and convective mixing can rapidly bring the Lithium from the photosphere to stellar interior. As a consequence Li is rapidly consumed during the PMS phase and its presence can be used as diagnostic for stellar youth \citep{RandichBPP2005}. However, the efficiency of Li depletion varies markedly with stellar mass. In stars more massive than $1\,M_{\odot}$, a radiative core develops during the contraction toward the Main Sequence, before the Li depletion is complete, reducing the efficiency of the convective mixing, and resulting in longer Li depletion timescales. In Sec. \ref{spec_sec} we classified the protostar \#5 as a K$2\pm1$ star, with $M_{\star} \geq 2.2\,M_{\odot}$. In this mass range, a timescale for Li depletion longer than $10\,Myrs$ can be expected \citep{JeffriesOliveira2005}. It is not surprising, then, that we observe a Li~$\lambda$6708 absorption line in protostar \#5.  \par

The best clue for the young age of both protostars is provided by the presence of circumstellar disks which are actively accreting, as well as the presence of outflow. In young clusters the fraction of members still bearing a disk is observed to decline from about 80\% in clusters younger than $1\,Myr$, to $\sim55\%$ at $3\,Myrs$ and $\sim15\%$ at $\sim5\,Myrs$ \citep{HaischLL2001, HernandezHMG2007, Mamajek2009}, indicating a rapid dissipation of disks. The timescale for disk dissipation is even faster in stars as massive as the protostars \#5 and \#7 (e.g. \citealp{LadaMLA2006}). The presence of a circumstellar disk, found by \citet{GuarcelloDWD2013} for protostar \#7 and by the present work for the protostar \#5, then, allows us to conclude that the central stars in these two proplyd-like objects are at most just a few Myrs old. Additionally, as we will show later, the spectra of both protostars show signatures of infall/outflow. The accretion process is expected to decline faster than the typical timescales for disk dissipation \citep{FedeleVHJ2010}. In conclusion, all the evidence points toward an age for these two objects ranging from a few $10^5\,yrs$ to $\sim2\,Myrs$. \par

\subsection{The circumstellar environment of protostar \#7}
\label{circum_sec}

\begin{figure}[]
        \centering
        \includegraphics[width=6cm]{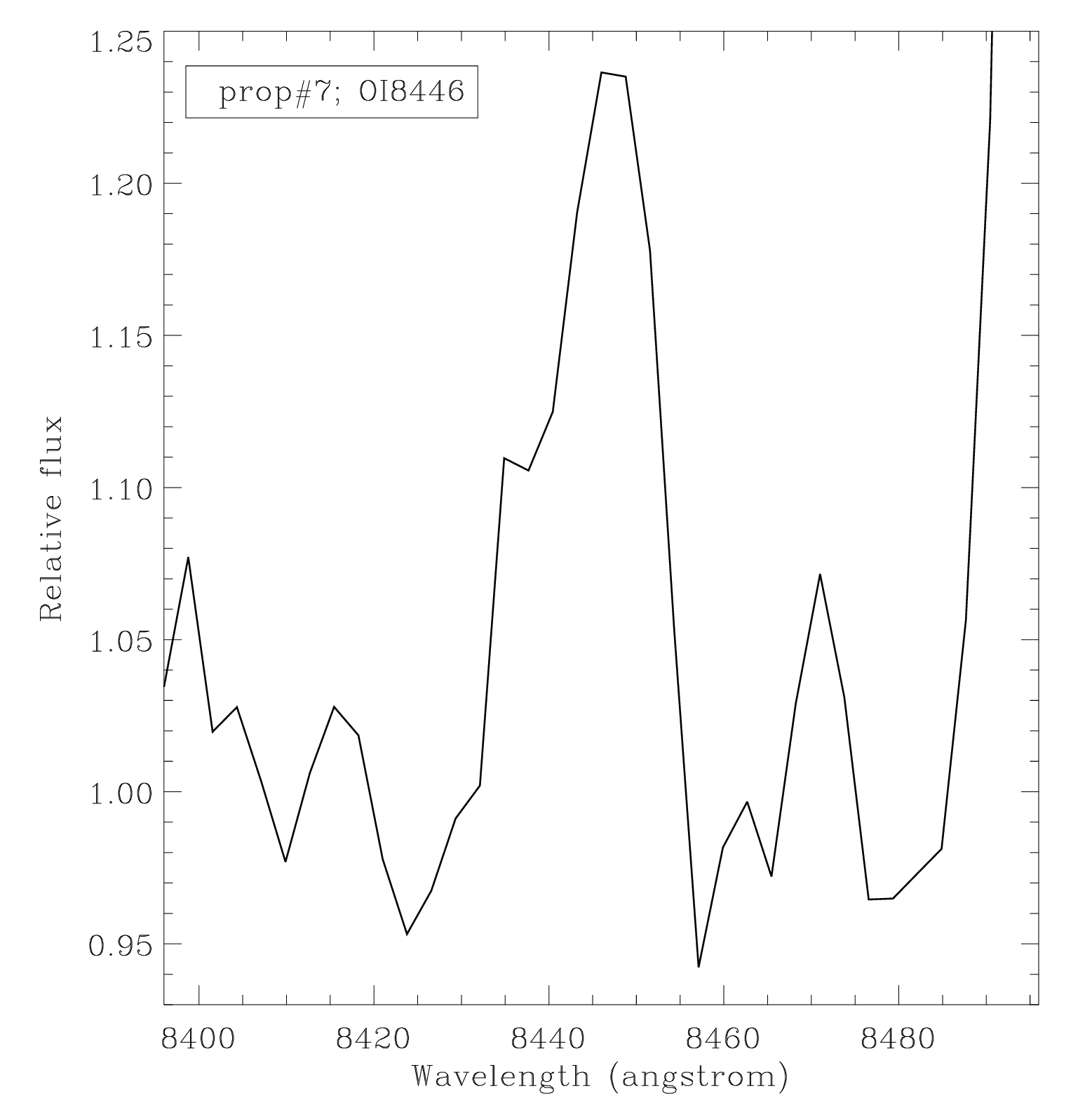}
        \caption{OI~$\lambda$8446 emission line of protostar \#7. The blue part of the CaI$\lambda$8498 emission line is also evident.}
        \label{OI8446_img}
        \end{figure}

One of the aims of our study is to understand whether the circumstellar disks around these two protostars are photoevaporating under the influence of the incident UV radiation. One clue in this sense is provided in the spectrum of protostar \#7 by the presence of the OI~$\lambda$8446 emission line (Fig. \ref{OI8446_img}). This line is produced primarily through the $Ly\beta$ fluorescence process \citep{BowenSwings1947}, in which the OI 3d $^3$D state is photoexcited upon absorption of $Ly\beta$ photons by the ground state, and then it decays with emission at 11287~\AA{}, 8446~\AA{} and 1304~\AA{}  in succession. This process was called {\it photoexcitation by accidental resonance} (PAR) by \citet{KastnerBhatia1995}, who constructed a detailed model for neutral oxygen and computed expected strengths of the various OI lines (also accounting for collisional excitation). The OI~$\lambda$8446 emission line has been observed in environments characterized by intense local UV fields, such as the  star forming regions in the Orion Nebula \citep{MuenchTaylor1974,Grandi1975} and the circumstellar environment of $\eta$ Carinae \citep{JohanssonLetokhov2005}, in Be stars \citep{MathewBSA2012} and in extragalactic objects such as Seyfert galaxies \citep{Grandi1980}. There is no diagnostic that allows us to connect the flux observed at OI~$\lambda$8446 with the intensity of the local $Ly\beta$ flux, but its presence in the spectrum indicates that the emitting gas is irradiated by UV photons. This UV flux can originate from the massive members of the association, if the intracluster extinction is not severe, or it can be produced by intense accretion, as discussed in Sect \ref{spec_sec}. \par

  The intense UV flux is expected to increase the level of ionization in the circumstellar gas. Evidence for the presence of a large amount of ionized gas in the circumstellar environment of protostar \#7 is provided by the low [OI]~$\lambda$6300/([SII] $\lambda$6716+[SII] $\lambda$6731) observed ratio. These lines will be discussed in detail later in this paper and used as diagnostics for electron density and mass loss rate. In this section, we compare the ratio of their fluxes with the values typically observed in other T~Tauri stars. In their analysis of the optical spectra of 42 T~Tauri stars, \citet{HartiganEG1995} found a typical [OI]~$\lambda$6300/([SII] $\lambda$6716+[SII] $\lambda$6731) ratio ranging between 2 and 4, with no stars with a ratio smaller than 1.5. It must be noted that these 42 T~Tauri stars are members of regions where no intense ionizing radiation field is expected, such as the Taurus Molecular Cloud. In the combined spectra of the protostar \#7, adopting an extinction $A_V=12^m$, the ratio is 1.38 (1.25 only in the OB4 spectra, where all the three lines are intense, while in the OB2 it was not possible to calculate this ratio), which is lower than what observed by \citet{HartiganEG1995}. Ratios similar to what we find in the protostar \#7 have been found in the proplyds in Orion \citep{BallySDJ1998,HenneyODell1999} and in the candidate photoevaporating star SO587 in $\sigma\,$Orionis \citep{RigliacoNRS2009}.\par

The presence of the OI~$\lambda$8446 emission line and the atypical [OI]~$\lambda$6300/([SII] $\lambda$6716+[SII] $\lambda$6731) ratio indicate that the  circumstellar environment of protostar \#7 is irradiated by an intense local UV field and that the fraction of ionized gas is larger than typically observed in T~Tauri stars, but similar to what is observed in photoevaporating disks.

\section{Accretion and outflow}
\label{acou_sec}
In this section we study the accretion and the outflow in the protostar \#7 using several emission lines. None of these, with the exception of H$\alpha$, are observed in the spectra of protostar \#5.

\subsection{Accretion rate and luminosity}
\label{accr_sec}

Both stars show signatures of active accretion, which is common in very young T~Tauri stars with disks. Typically the mass accretion rate observed in T~Tauri stars declines with the age of the disk (e.g. \citealp{HartmannCGD1998}) and it varies with the mass of the central star with an empirical scaling law $\dot{M} \propto M^2_{\star}$ \citep{AlexanderArmitage2006}. One common signature for active accretion is the presence in the spectra of the 8498~\AA{}, 8542~\AA{}, and 8662~\AA{}  Ca~II triplet lines \citep{HerbigSoderblom1980,MuzerolleHC1998}, which can be easily excited by the veiling continuum produced in the accretion spots. Several clues suggest that the Ca~II triplet is emitted by gas in the inner boundary of circumstellar disks and in the accretion flow \citep{KwanFischer2011}. \par

\begin{figure}[]
        \centering
        \includegraphics[width=6cm]{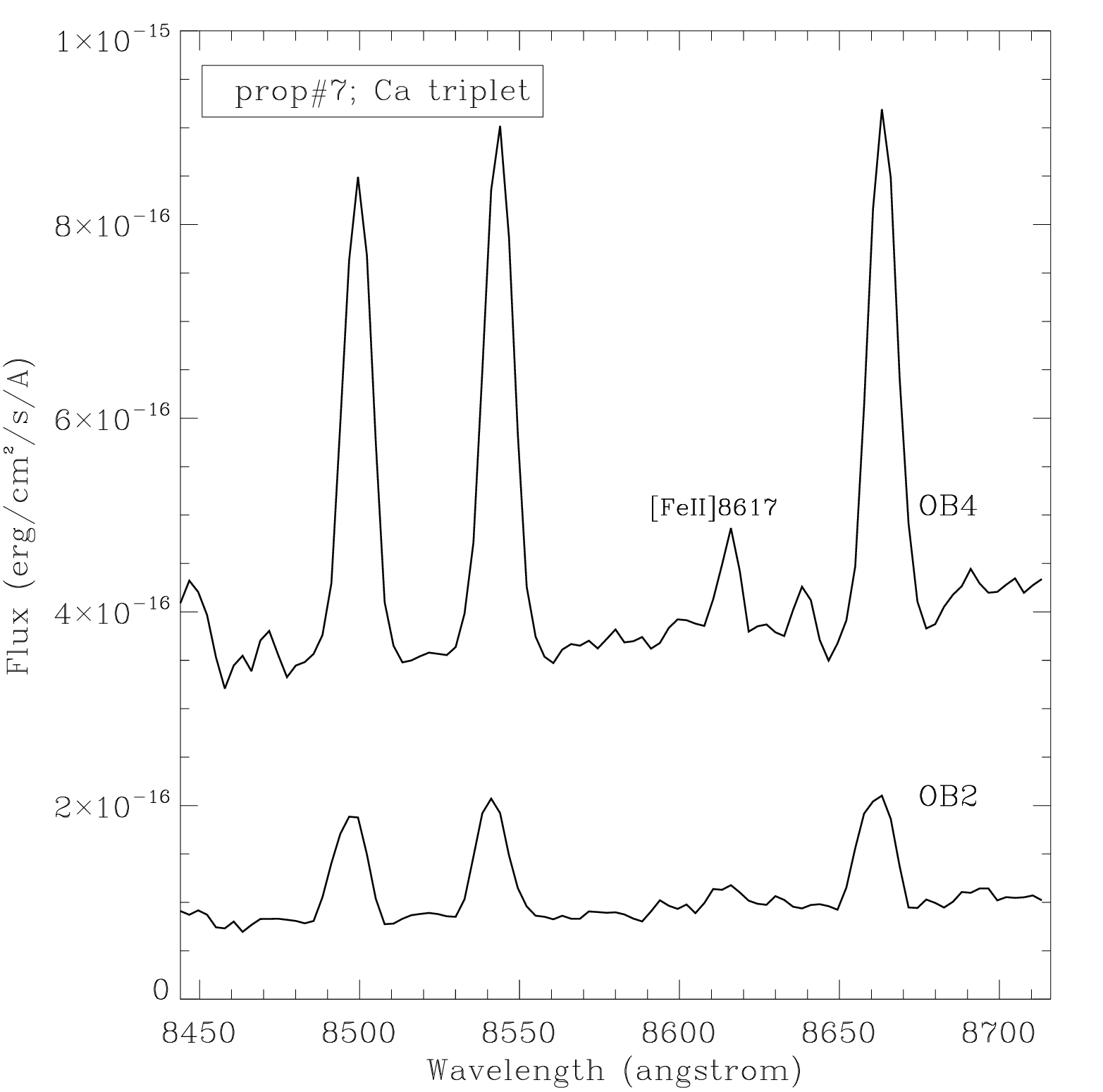}
        \caption{Region of the spectrum of protostar \#7 containing the infrared Ca~II triplet and the [Fe~II]~$\lambda$8617 emission line, observed in OB2 (fainter spectrum) and OB4 (brighter spectrum). The offset between the two spectra is due to the variability of this source}
        \label{CaT_img}
        \end{figure}

  The red part (i.e. $\lambda \geq 8400\,$\AA{}) of the protostar \#7 spectrum is dominated by an intense Ca~II triplet (see Fig. \ref{CaT_img}). The intensity ratio between the three lines provides clues about their nature: If emitted by optically thin material their ratio is given by the ratio of their $gf$-values, 1:9:5 for 8498:8542:8662~\AA{}, respectively \citep{WieseSmithMiles1969}; while if emitted by optically thick gas, the lines saturate and the ratio flattens out. The ratio we observe in the combined spectrum of the protostar \#7 is 1:1.1:1, with slight differences between the OB2 and OB4 spectra. This ratio indicates that the Ca~II triplet is emitted by optically thick material, as typically observed in accreting T~Tauri stars. \par

A correlation between the intensity of the Ca~II~$\lambda$8542 line and the accretion luminosity was found by \citet{MuzerolleHC1998} and \citet{CalvetHartmannStroom2000}:

\begin{dmath}
log\left(L_{acc}/L_{\odot} \right) = \left(0.85 \pm 0.12 \right) log\left(L_{8542}/L_{\odot} \right) +  \newline \left( 2.46 \pm 0.46 \right)
\label{acr_eq}
\end{dmath}

After the reddening correction (with $A_V=12^m$), the flux emitted in the Ca~II~$\lambda$8542 line by protostar \#7 is $4.72\times10^{-13}\,$erg/cm$^2$/s/\AA{} in the OB2 spectrum and $1.98\times10^{-12}\,$erg/cm$^2$/s/\AA{} in OB4. Converting these values into luminosity, and using equation \ref{acr_eq}, these fluxes correspond to $L_{acc}=5.45\times10^{34}\,erg/s$ (OB2) and $L_{acc}=1.84\times10^{35}\,erg/s$ (OB4). This difference in flux observed in the two OBs for the Ca~II~$\lambda$8542, as well as other in other lines as described later, is intrinsic given that the spectra have been flux-calibrated (Sec. \ref{data_sec}). \par

  To convert $L_{acc}$ into mass accretion rate, we use the equation defined by \citet{GullbringHBC1998}:

\begin{equation}
\dot{M}=\frac{L_{acc}R_{\star}}{GM_{\star}}\left(1-\frac{R_{\star}}{R_{inner}} \right)^{-1}
\label{acrate_eq}
\end{equation}

Since the spectral classification of the protostar \#7 is not accurate, we consider a range of possible spectral classes, and we calculate the corresponding $\dot{M}$ using equation \ref{acrate_eq} and the stellar parameters in this range from the $0.5\,Myr$  isochrone of \citet{SiessDF2000}: we adopted $4\,M_{\odot}<M_{\star}<5\,M_{\odot}$ and $8\,R_{\odot}<R_{\star}<10.4\,R_{\odot}$. Using these values, the accretion rate ranges between $1.13 - 1.17\times10^{-6}\,$M$_{\odot}$/yr for OB2, and $3.8 - 4.0\times10^{-6}\,$M$_{\odot}$/yr for OB4. \par

\begin{figure*}[]
        \centering
        \includegraphics[width=14cm]{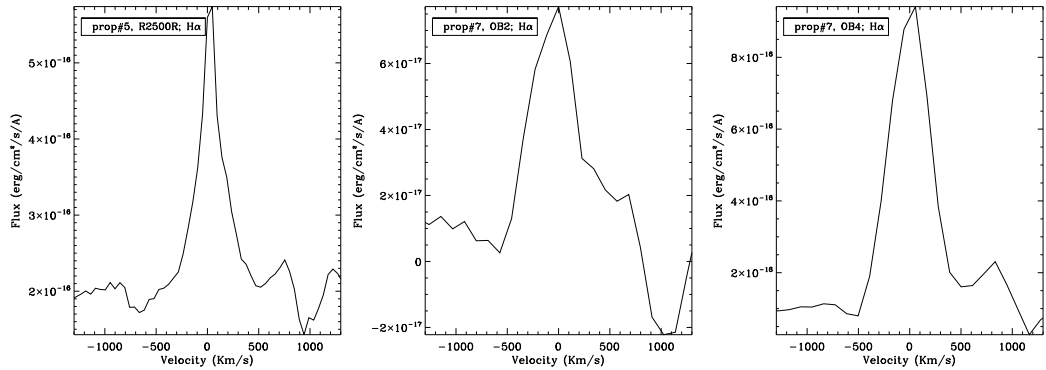}
        \caption{$H\alpha$ emission line observed in the R2500R spectrum of protostar \#5 and the spectra of protostar \#7 in two OBs, with the redshifted and blueshifted (at about $\pm 500\,$km/s) absorption features.}
        \label{Ha_img}
        \end{figure*}

  Another line usually observed in accreting T~Tauri stars which can be used as a diagnostic of mass accretion rate is the $H\alpha$ line. For instance, \citet{NattaTMR2004} correlated the H$\alpha$ 10\% width with the accretion rates observed in a sample of 19 very low mass objects in Chamaleon I and $\rho$ Ophiuchi. The use of the width at 10\% of the peak is meant to reduce the sensitivity to contamination from the nebula emission around the star, which is not always trivial to remove. In these two objects the nebula contribution is very intense, and it arises from the intense emission in H$\alpha$ from the envelope \citep{WrightDDG2012}, so we have not attempted this estimate. Fig. \ref{Ha_img} shows the H$\alpha$ profile observed in the R2500R spectrum of protostar \#5, and that in the protostar \#7 observed in the two OBs. In all the profiles, the base of the line is broad, as expected in accreting objects, with the narrow peak due to nebular emission being more evident in \#5. Other features include the evident absorption at $\pm 500\,$km/s (more evident in \#5). Red-shifted absorption features at velocities exceeding $100\,$km/s are a signature of accretion \citep{AppenzellerWolf1977,EdwardsHGA1994}, while blue-shifted absorption, more rarely observed in T~Tauri stars, indicates the presence of outflows \citep{HartiganEG1995}. The H$\alpha$ profiles in the two targets, then, even if not useful to quantify the mass accretion rate, provide evidence for presence of infall and outflow in both objects. \par 

Another clue about the presence of intense accretion in the protostar \#5 is provided by the IPHAS colors. Using the IPHAS colors of normal stars derived by \citet{DrewGIA2005}, a K2$\pm1$ star with r-i=$2.2^m$ is supposed to have r-H$_{alpha}$=0.66$^m$, while this protostar has r-H$\alpha$=1.63$^m$. This is significantly larger than the color expected for normal stars, and compatible with an intense H$\alpha$ emission due to accretion \citep{BarentsenVDG2011}, although some level of contamination of the H$\alpha$ photometry of this object may be present.

\subsection{Outflow}
\label{out_sec}

   Classical T~Tauri stars often show signatures of outflow. The presence of a slow and wide wind driven off from the circumstellar disk is used to explain both the blueshifted forbidden emission lines and redshifted absorption features which are associated with the emission lines, which are typically observed in T~Tauri star, with the hypothesis that the disk obscures the part of the wind which is moving away from the observer. The presence of high-velocity collimated jets has also been proved by direct high-resolution imaging observations of T~Tauri stars such as DG~Tau and HL~Tau (i.e. \citealp{Solf1989}), and the presence of blue-shifted high-velocity components in the emission lines \citep{HartiganEG1995}. There is a growing body of evidence that these outflows are powered by the accretion in the disk, and a correlation between the mass loss rate and mass accretion rate is observed, with the former usually being two orders of magnitude less than the latter \citep{CabritESS1990}. \par
  
In the spectra of the protostar \#7 several Forbidden Emission Lines (FELs) are observed. FELs, such as [OI]~$\lambda$6300 or the [SII] $\lambda$6731+6716 doublet, are observed from a variety of low-density astrophysical environments, where the decay of electrons from meta-stable states is more likely to occur spontaneously rather than after collisions. In young stellar objects, they usually serve as powerful tracers and diagnostics for outflow activity close to star and disk \citep{Hirth1994}. Fig. \ref{FEL_img} shows the main FELs observed in the spectra of protostar \#7. The upper panel shows the [SII] doublet, used as a diagnostic for electron density. We observe a significant variation of both the shape and the intensity of this doublet among the two OBs. In OB2 it is almost undetected, with an absorption feature that may not be real; in OB4 the two lines of the doublet are evident and easily distinguishable. The second panel shows the region of the spectra containing the [OI]~$\lambda$6300 and [OI]~$\lambda$6363 lines. The former is present in both spectra with an asymmetric shape that will be discussed later; the latter is present only in the OB4 spectrum. The third panel shows the [OI]~$\lambda$5577 line. The OB2 spectrum is very noisy and it is not clear whether the line is present or not, while in OB4 the line is more evident. Finally, the fourth panel shows the [Fe~II]~$\lambda$7155 line, which is present and asymmetric in both spectra. The [Fe~II] FEL at 8617\AA{} is observed and shown in Fig. \ref{CaT_img}. \par

\subsection{Profiles of the FELs}
\label{prof_sec}

   The centroid of the FELs observed in T~Tauri stars is always blueshifted. This is an indication that the source of the emission is not the star itself, but outflowing material in the wind and/or a jet \citep{CabritESS1990}. Another important feature of the FELs is that they are often multi-component (i.e., \citealp{HartiganEG1995, HirthMS1997, WhelanRD2004}), with a superposition of a low-velocity (typically few km/s) and a high-velocity (typically a few hundreds km/s) component (LVC and HVC, respectively). Strong evidence exists that the HVC observed in FELs is emitted by stellar jets, such as images at high spatial resolution where it is possible to resolve the region where the line is actually produced (i.e. \citealp{KepnerHYS1993,WhelanRD2004}), while the LVC is emitted by a broad and slow disk wind \citep{KwanTademaru1988,Kwan1997}. \par

The FELs observed in protostar \#7 are blueshifted as reported in Table \ref{shift_tb}, and the two velocity components can be distinguished only in few lines, given the low resolution of our spectra. In the few cases where two peaks are present, Table \ref{shift_tb} lists the centroid velocities of the low velocity component. The velocities are larger than the values typically observed in T~Tauri stars (i.e. \citealp{CabritESS1990}), but examples of stars with such high-speed outflows are present in the literature (i.e. DG Tau and LkH$\alpha\,$321, \citealp{WhelanRD2004}). \par

  The centroids of the FELs were measured fitting the line profile as explained in Sect. \ref{data_sec} using \emph{SPLOT}. Given that the results of the fit change slightly when adopting different starting points across the line profile, we repeated the fit several times for each line changing the starting point each time. The results listed in Table \ref{shift_tb} are taken from a median of the measured values. The errors listed are whichever are the larger among the error provided by the line profile fit or the one representing the range in which the nominal value of the centroid varied. We note that our centroid velocities in Table \ref{shift_tb} should be interpreted with caution: in the R1000R low-resolution spectra the LVC and HVC are usually not distinguishable, and we have no means of determining an accurate correction for the radial velocity of protostar \#7. \par

The only FELs in which the HVC and LVC are distinguishable are the [OI]~$\lambda$5577 and [OI]~$\lambda$6363 lines. After deblending the two components of these two lines using \emph{SPLOT}, the LVCs were observed respectively at $-93\pm53\,$km/s and $-150\pm23\,$km/s, while the HVCs were at $-296\pm17\,$km/s and $-299\pm27\,$km/s, respectively. The presence of FELs, then, demonstrate that protostar \#7 is characterized by intense outflow, with marginal evidence also for the presence of a stellar jet. \par

\subsection{Electron density in the outflow of protostar \#7}
\label{dens_sec}

An excellent diagnostic for the electron density is provided by the [SII] $\lambda$6716/[SII] $\lambda$6731 ratio, which is independent from the temperature of the emitting gas \citep{CzyzakKA1986}. In T~Tauri stars, most of the observed [SII] ratios are close to the saturation value, which occurs at $N_e \geq 5\times10^3\,$cm$^{-3}$ (i.e. \citealp{Hamann1994}). \par
  
  The OB4 spectra of protostar \#7, shown in Fig. \ref{FEL_img}, have a clear and intense [SII] doublet. In OB2, the [SII] $\lambda$6731 line is still evident, but the spectrum is very noisy. To calculate the [SII] $\lambda$6716/[SII] $\lambda$6731 ratio in the combined OB4 spectrum, we deblended the two lines using \emph{SPLOT} in the dereddened spectra. The best fit is obtained with two Gaussian profiles centered at 6711\AA{}  and 6725\AA{}, with a flux ratio of 0.47, which is close to the saturated regime \citep{CzyzakKA1986}. An accurate estimate of the electron density in the outflowing material of protostar \#7 can be obtained with the relation found by \citet{ProxaufOK2014}, which gives $log \left(N_e \right)=4.03\,$cm$^{-3}$. The symmetry of the two lines in the [SII] doublet does not allow us to attribute this electron density to the stellar wind or to an eventual stellar jet, which can be reasonably expected to have different temperature and density \citep{HartiganEG1995}. Lacking spatially resolved data for protostar \#7 and any evidence for a HVC in these two lines, this electron density will be used as an average value for the whole stellar outflow hereafter. \par 

\subsection{Mass loss rate in protostar \#7} 

\begin{figure*}[]
        \centering
        \includegraphics[width=6cm]{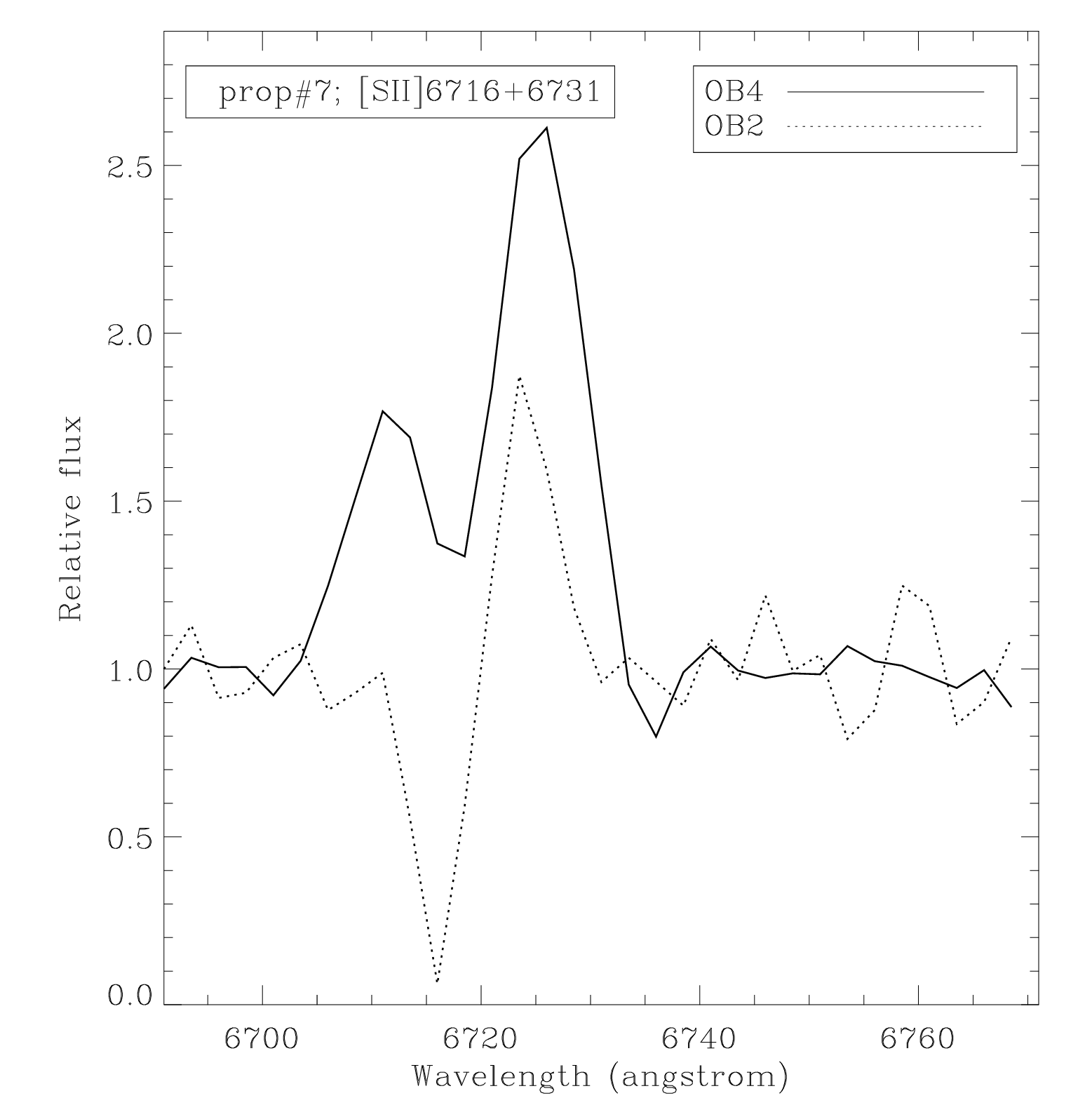}
        \includegraphics[width=6cm]{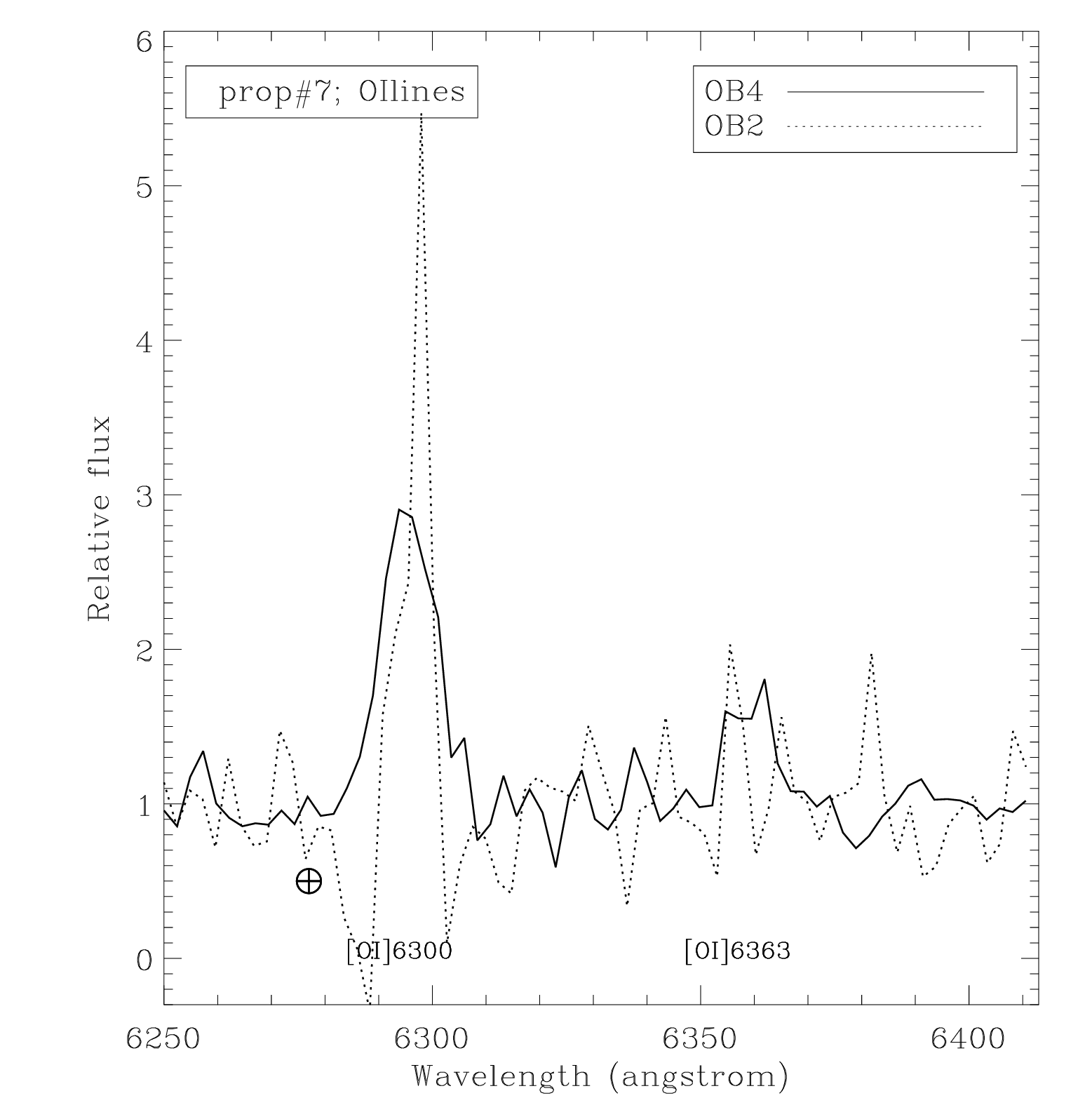}
        \par
        \includegraphics[width=6cm]{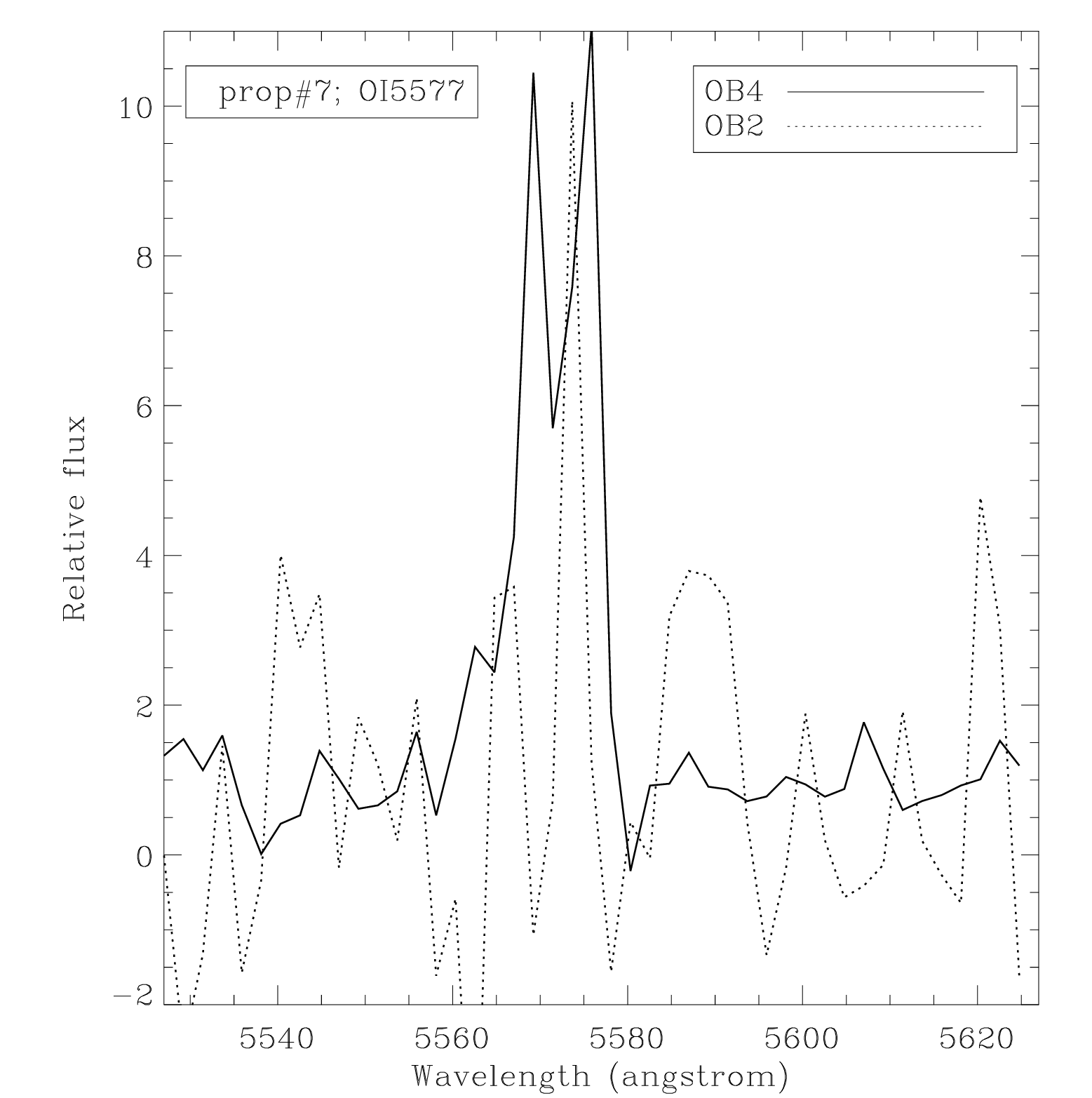}
        \includegraphics[width=6cm]{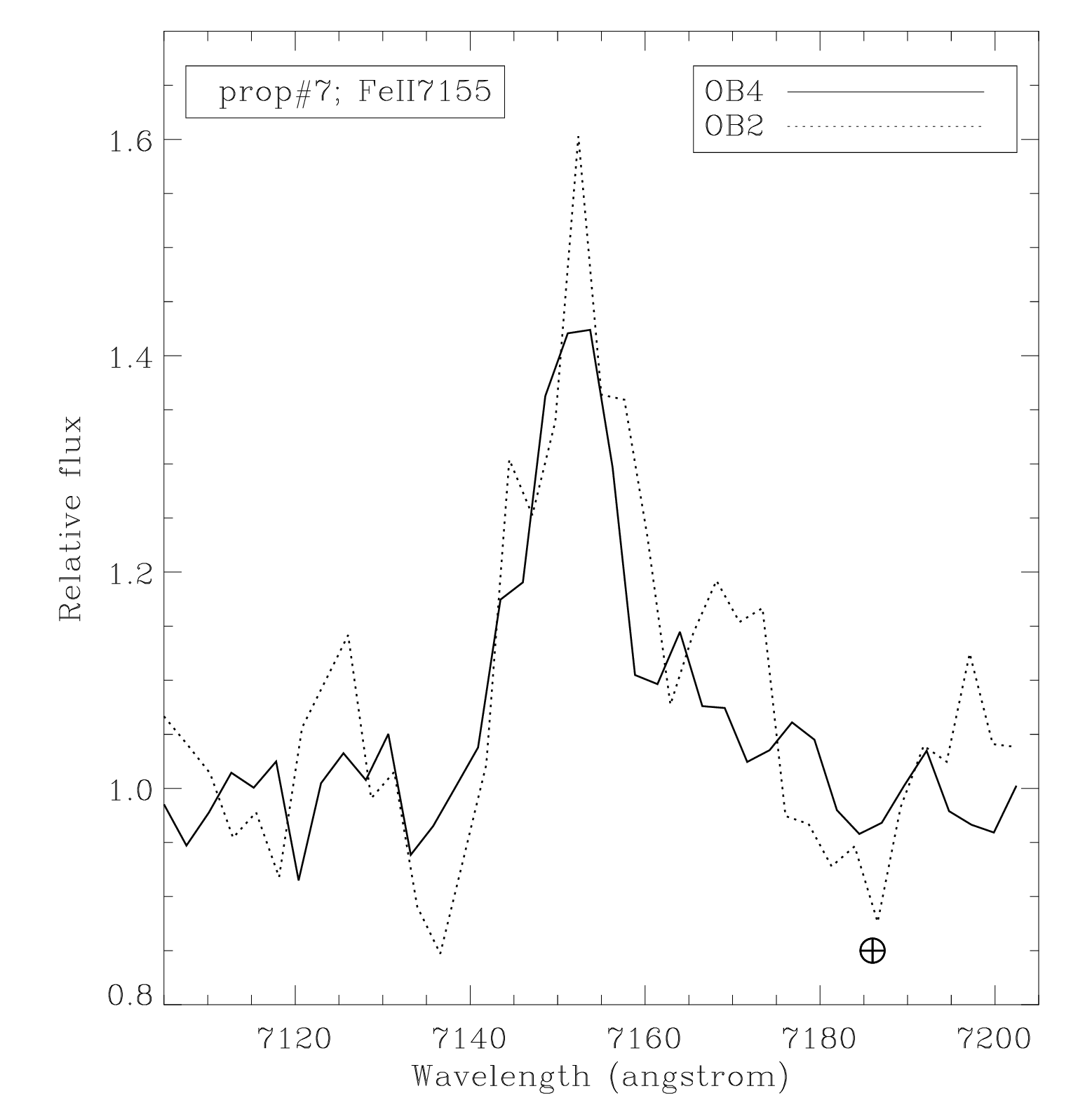}
        \caption{Forbidden emission lines observed in the spectra of protostar \#7. In all the panels, the solid lines mark the spectrum taken during OB4, the dotted during OB2. The panels show: [SII]  $\lambda$6716+ $\lambda$6731 doublet (upper left), the [OI]~$\lambda$6300 and [OI]~$\lambda$6363 lines (upper right), the [OI]~$\lambda$5577 line (lower left), and the [Fe~II]~$\lambda$7155 line (lower right). The positions of telluric absorption features are also indicated.}
        \label{FEL_img}
        \end{figure*}

An intense [OI]~$\lambda$6300 emission line from protostar \#7 was seen in both OBs (See Fig. \ref{FEL_img}). The mass loss rate due to the outflow can be calculated using the equations found by \citet{HartiganEG1995}:

\begin{dmath}
\dot{M}=2.27\times10^{-10}\left(1+\frac{N_c}{N_e}\right)\left(\frac{L_{[OI] \lambda6300}}{10^{-4}L_{\odot}}\right) \times \left(\frac{V_{\perp}}{150\,km/s}\right)\left(\frac{l_{\perp}}{2\times10^{15}\,cm}\right)^{-1}\,M_{\odot}/yr
\label{mdot_1_eq}
\end{dmath}

We adopted the value of $150\,$km/s for the outflow velocity $V_{\perp}$, and $1.97\times10^{6}\,$cm$^{-3}$ for the critical density $N_c$, as in \citet{HartiganEG1995}; we also used the electron density $N_e$ found using the [SII] doublet ratio (Sect. \ref{dens_sec}) and the slit width $l_{\perp}$ of $1.2^{\prime\prime}$. We calculated $L_{[OI] \lambda6300}$ from the line flux observed in the dereddened spectra (adopting A$_V=12^m$) and adopting the distance of $1400\,$pc \citep{RyglBSM2012}. Using the OB4 spectra (from which we have been able to calculate the electron density of the outflow), Eq. \ref{mdot_1_eq} gives a mass loss rate of $5.0\times10^{-5}\,$M$_{\odot}$/yr. In this calculation we ignored the sky contamination of the [OI]~$\lambda$6300 emission line. This is justified since the intensity of the sky line is negligible with respect to the intense emission from the protostar\footnote{See the typical sky spectrum from GTC at http://www.gtc.iac.es/observing/toolbox.php}. Additionally, the sky line is centered on $6302\,$\AA{}, while the emission from the protostar is blueshifted and well distinguished from this contamination (see Fig. \ref{FEL_img} and Sect. \ref{prof_sec}). \par

  It is also possible to estimate the mass loss rate from the [SII] $\lambda$6731 line intensity \citep{HartiganEG1995}:

\begin{dmath}
\dot{M}=3.38\times10^{-8}\left(\frac{L_{[SII] \lambda6731}}{10^{-4}L_{\odot}}\right)\left(\frac{V_{\perp}}{150\,km/s}\right) \times \left(\frac{l_{\perp}}{2\times10^{15}\,cm}\right)^{-1}\,M_{\odot}/yr
\label{mdot_2_eq}
\end{dmath}

Using this equation, we obtain a mass loss rate of $2.2\times10^{-5}\,$M$_{\odot}$/yr, similar to the value calculated from Eq. \ref{mdot_1_eq}. 

\section{Variability}
\label{varia_sec}

T~Tauri stars are known to be variable sources. For instance, in the Orion Molecular Cloud 70\% of the members with disks and 44\% of those with no disks are observed to be variable in the infrared (the YSOVAR survey, \citealp{Morales-CalderonSHG2011}), and variability in young stellar objects is observed in recent NIR data \citep{ContrerasPenaLFK2014}. This variability can be due to several phenomena, such as: slow or impulsive variation of the mass accretion rate or of the geometry of the accretion flow; variable shadowing by circumstellar material, which can be associated with a warped disk; stellar rotation with a consequent modulation of the emission observed from the accretion spots which are non uniformly distributed on the stellar surface; flares arising from magnetic activity (see, for instance, \citealp{CodySBM2014}). \par

  OB2 and OB4 observations of protostar \#7 were taken 4 days apart (see Table \ref{log_tb}), and significant variability has been observed through comparison of the two epochs, with a flux variation of a factor between 2 and 3 in the two OBs. This variability can be compared with the results in Orion from the YSOVAR survey, where the typical peak-to-peak changes in the infrared is about $0.2^m$, though with some more extreme cases showing a variability of $1.8^m$. Converted into flux unity, these values correspond to variations of a factor 1.2 and 5.2, respectively, similar to what we observed in protostar \#7. \par

The fact that this variability involves not only the continuum from the central object, but also all the emission lines (see Table \ref{flux_tb}) indicates that it is mainly due to changes in the intensity of the infall/outflow activity. For instance, the CaT lines are more intense in OB4 by a factor $\sim4$ compared with OB2, corresponding to an increase in the accretion rate of a factor 3.4. The increase in intensity of all the FELs from OB2 to OB4 indicates that the larger accretion rate in the second day of observation is quickly accompanied by an increase in the intensity of the outflow. Unfortunately, the lack of a detection of the [SII]~$\lambda$6717 line in the OB2 spectrum does not allow us to estimate the electron density. However, adopting the same $N_e$ found in OB4, we obtain from the OB2 spectra $\dot{M}_{wind}\sim5\times10^{-6}\,$M$_{\odot}$/yr, which is one order of magnitude smaller than the value observed during the OB4. Variations of similar magnitude have been reported in  the literature, such as in the T~Tauri star XZ Tau, which showed changes in the intensity of the [OI]~$\lambda$6300 line by a factor 2.6 in two months \citep{ChouTMB2013}. Variations in the FELs are not the same for all the lines (i.e. \citealp{HamannPersson1992}), suggesting that the observed variability results from a combination of factors. \par

  More surprising is the variability observed in protostar \#5. {\bf The region around $7500\,$\AA{} is common in all the three spectra of this source. After dereddening the spectra adopting A$_V=7.5^m$ (calculated for a 1 Myr old K2 star at the distance of Cyg~OB2 using the isochrones from \citealp{SiessDF2000}), the flux at $7500\,$\AA{} in the R100B spectrum is $\sim9\times10^{-15}\,$erg/cm$^2$/s/\AA{}, $\sim7\times10^{-14}\,$erg/cm$^2$/s/\AA{} in the R2500R and $\sim1.5\times10^{-16}\,$erg/cm$^2$/s/\AA{} in the R2500I spectrum. Such variation is greater than the typical variation observed in YSO in recent surveys, such as the Coordinated Synoptic Investigation of NGC2264, even in bursting stars \citep{CodySBM2014, StaufferCBA2014}. Rapid variations of several magnitudes (the flux variation between the two R2500 spectra roughly corresponds to a magnitude variation of $6.6^m$) are usually observed in the FU Orionis stars \citep{Herbig1966} as a consequence of rapid bursts of mass accretion. This hypothesis is hardly supported, however, by the fact that the decay observed in such stars is longer than the time difference between the two R2500I and R2500R observations (slightly larger than 30 minutes) and by the lack of evidence of a very high mass accretion rate (during the burst in the Fu Ori stars the mass accretion rate increases up to $\sim10^{-4}\,$M$_{\odot}/$year) in the observed spectra. Further observations are required in order to confirm the peculiar variability of this source and rule out the possibility that it can be due to some calibration issue.}

\section{Discussion}
\label{discussion}

  The proplyd-like objects near Cygnus~OB2 are unusual and important sources. Evaporating proplyds have already been observed in other young clusters hosting massive stars. Those which have been best characterized are those in the Trapezium in Orion, at distances $d<0.5\,$pc from the O6V star $\Theta^1\,$Ori \citep{BallySDJ1998}. The proplyd-like objects near Cyg~OB2 have three peculiar characteristics with respect those in the Trapezium: their dimension (ranging between $18000\,$AU and $113000\,$AU, while those in Orion are between $40\,$AU and $400\,$AU); their distance from the ionizing sources ($6-14\,$pc) and the fact that some of them are thought to host an intermediately massive star in their center \citep{ComeronPRS2002}. \par

With the present work, we obtained some insight into the nature of two of the proplyd-like objects in the area surrounding Cygnus~OB2: \#5 and \#7 \citep{WrightDDG2012}. In particular, the central star in object \#7 is supposed to be a G star younger than $1\,Myrs$, still in its class~I phase \citep{GuarcelloDWD2013}; while protostar \#5 contains a K$2\pm1$ star of similar age. In the optical spectra analyzed here, we found evidence of active accretion in both stars, and both an outflow (with marginal indications for the presence of a stellar jet) and strong variability. \par

  In the case of protostar \#7, we derived a mass loss rate $\sim5\times10^{-6}\,$M$_{\odot}$/yr in OB2 and $\sim5\times10^{-5}\,$M$_{\odot}$/yr in OB4, and mass accretion rate of $1.13 - 1.17\times10^{-6}\,$M$_{\odot}$/yr and $3.8 - 4.0\times10^{-6}\,$M$_{\odot}$/yr in OB2 and OB4, respectively. The mass loss rate is larger than the mass accretion rates, suggesting that the mass evolution in this disk is dominated by the outflow. This is not a common situation in T~Tauri stars. For instance, in the sample of 42 T~Tauri stars with outflows studied by \citet{HartiganEG1995}, the mass loss rate are at least one order of magnitude less intense than the mass accretion rate, with the only exception being HN Tau, whose mass loss rate was 0.4 times the mass accretion rate. Other studies, such as \citet{CabritESS1990}, indicate that in T Tauri stars typically the mass loss rate is $0.1 - 0.01$ times the mass accretion rate. \par

The mass loss rate in protostar \#7 is particularly intense. In their spectroscopic study of the proplyds in the Orion Nebula Cluster \citet{HenneyArthur1998} and \citet{HenneyODell1999} estimated that the mass loss rate of 31 proplyds ranges between $2.5\times 10^{-8}\,$M$_{\odot}$/yr and $1.6\times 10^{-6}\,$M$_{\odot}$/yr. This upper limit is about 30 times smaller than the mass loss rate observed in the object \#7 during OB4. The absolute values of mass loss rate and mass accretion rate that we derived must, however, be used with caution. The line fluxes have been taken after dereddening the spectra adopting $A_V=12^m$. This extinction is reasonable if the central protostar is a G star, but this classification is mainly based on the lack of the Paschen series and TiO bands, rather than the presence of absorption lines typical of these spectral types. Adopting a different spectral type means that a different value of extinction and stellar parameters in Eq. \ref{acrate_eq} must be used. To verify whether the results of this study are qualitatively consistent and not affected by the uncertainty on stellar extinction, we calculated the mass loss rate, mass accretion rate, and [OI]~$\lambda$6300/([SII] $\lambda$6716+[SII] $\lambda$6731) ratio adopting different values of extinction: $A_V=0^m,\, 6^m,\,8^m$. The [SII] $\lambda$6716/[SII] $\lambda$6731 ratio is independent from the adopted extinction. The results are shown in Table \ref{ext_tb}. The numerical values of mass loss rate and mass accretion rate change adopting different extinctions, but the mass loss rate always exceeds the mass accretion rate. \par  
  
  In the calculation of the mass loss rate and mass accretion rate we have not taken into account the uncertainty on flux measurement, which is not a trivial task. The main source of flux uncertainty is the fraction of stellar flux which is lost during the observation because of the slit width. This loss is compensated by the flux calibration described in Sect. \ref{data_sec}. To verify whether a less than optimal calibration could affect our result, we can consider the worst case. If the calibration was completely ineffective, given that the upper limit of seeing during the nights of observation is comparable with the slit width (see Table \ref{log_tb}) the fraction of stellar flux lost during the observation should be about 50\%. Increasing the flux observed in [OI]~$\lambda$6300 and CaII~$\lambda$8542 by this fraction, the mass accretion rate is always smaller than the mass loss rate. We conclude then that our results are not affected by flux calibration uncertainties. Assuming that the flux calibration is reliable, the best estimate on the uncertainty on the flux value can be derived from the signal to noise ratio in the region close to the two lines we used. The OB2 spectra are more noisy than the OB4 ones, and in both OBs the blue part of the spectrum is more noisy than the red part. For [OI]~$\lambda$6300, an uncertainty between 33\% and 50\% can be expected on the flux estimate from OB2, and between 14\% and 20\% in OB4. For the CaII~$\lambda$8542 line the expected uncertainties range from 10\% to 14\% (OB2) and are $\sim$6\% in OB4. Considering these errors and repeating the calculation of the accretion rate and mass loss rate in the two OBs, we find that our estimate of the mass loss rate suffers uncertainty of 40\% in OB2 and 16\% in OB4, while the estimate of mass accretion rate is uncertain by 8.3\% in OB2 and 5.1\% in OB4. \par 
  
One hypothesis concerning the nature of these objects is that they are stars with photoevaporating disks \citep{WrightDDG2012}. While in normal T Tauri stars the mass loss rate is always observed to be about two orders of magnitude less intense than the mass accretion rate, photoevaporating disks are observed with mass loss rates more intense than the mass accretion rates, as observed in $\sigma\,$ Orionis \citep{RigliacoNRS2009} and the Trapezium \citep{HenneyODell1999}. This characteristic is shared by protostar \#7, supporting the hypothesis that the disk in this star is photoevaporating. Furthermore, the OI~$\lambda$8446 emission line demonstrates the presence of an intense Lyman continuum in the circumstellar environment of \#7, and the low [OI]~$\lambda$6300/([SII] $\lambda$6716+[SII] $\lambda$6731) ratio, with a value slightly larger than those typically observed in the ionization fronts of proplyds \citep{BallySDJ1998,StorzerHollenbach1999}, indicates that the environment is characterized by a high level of ionization. \par
  
Our data, then, provide strong evidence that the disk in protostar \#7 is photoevaporating. However, they do not allow us to differentiate between external or internal driving. The photoevaporation can be externally induced by the UV radiation emitted by the OB stars in the center of Cygnus~OB2, which also sculpts the envelope into the typical cometary shape observed in most of the Orion proplyds \citep{BallySDJ1998}. However, the UV radiation field can also be produced by active accretion onto the central star \citep{GullbringHBC1998} and the photoevaporation can in this case be ``self'' induced \citep{AlexanderCP2006}. Without considering the absorption of the UV flux from the intracluster material, the FUV and EUV fluxes emitted by the O stars in Cygnus~OB2 are intense enough to induce the observed photoevaporation of the disk in protostar \#7. As explained, however, a significant absorption of the EUV photons is expected, mainly by the ionization front in the envelope, while FUV photons can reach and irradiate the disk more easily. To obtain a more quantitative estimate of the absorption of EUV photons, we can combine equations 5 and 9 in \citet{JohnstoneHB1998} and calculate the ionization radius as:

\begin{equation}
r_{if}=\left( \frac{\dot{M}_{loss}}{12\pi m_H c} \right)^{2/3}\left( \frac{\alpha d^2}{F_{EUV}} \right)^{1/3}
\label{rif_eq}
\end{equation}

where $m_H$ is the mean particle mass per hydrogen atom; $c\sim10\,$km/s is the velocity at the ionization radius; $\alpha=2.6\times 10^{-13}\,cm^3\, s^{-1}$ is the recombination coefficient for hydrogen at $10^4 K$; $d$ is the distance from the ionizing sources and $F_{EUV}=5.6\times10^{50}\,$photons/s is the total emitted EUV flux (summing the emission from all the O stars). Equation \ref{rif_eq} gives an ionization radius of $5903\,$AU ($4.2^{\prime\prime}$), which is similar to the distance between the central star and the bright ionization front visible in the IPHAS image (the distance can be understood as an effect of the intracluster absorption of the EUV radiation, that we ignored here). IPHAS images of Cyg~OB2 \citep{Vink2008,GuarcelloDWD2013} show that some residual intracluster gas is still present in the region. This gas can absorb part of the EUV radiation emitted by the O stars in the center of the association before reaching the proplyd-like objects. For instance, an extinction of $A_V=1.38^m$, which is easy to achieve, is capable of reducing the EUV flux by a factor 100 (calculated from the extinction law of \citealp{CardelliCM1989}). The presence of an ionization front around the central protostar in object \#7 and the possibility that part of the EUV radiation is extinguished by the intracluster gas, suggest that the UV radiation created by the accretion process must play an important role in the photoevaporation of the circumstellar disk in this protostar. This is also suggested by the observation of a stronger [OI]~$\lambda$8446 line during OB4, when the accretion rate was higher. In both scenarios, however, the forbidden emission lines can be produced by the evaporating flow, which is mainly neutral before reaching the ionization front where the EUV radiation is absorbed \citep{StorzerHollenbach1999}. \par


  In theory, part of the [OI]~$\lambda$6300 emission could arise from the disk itself. The proplyds in Orion are in fact also observed as bright silhouettes in [OI]~$\lambda$6300 \citep{BallySDJ1998}. \citet{StorzerHollenbach1998} have shown that this emission can be produced by the photodissociation front associated with the disks in these photoevaporative systems. This line, in fact, can be excited thermally by collisions if the temperature of the gas in the disk is $T>3000\,K$, which is possible when the disk is illuminated by FUV radiation with an intensity of $10^3-10^4\,$G$_0$. Also a non-thermal component due to the dissociation of OH molecules can be present \citep{StorzerHollenbach2000}. However, the emission from the disk should be symmetrical, broadened by the keplerian velocity of the disk, and centered on $\lambda=6300$\AA{}. We do not observe such a broadened component in the [OI]~$\lambda$6300 profile in the spectra of protostar \#7, suggesting that it is absent or much weaker than the blueshifted component due to the outflow. \par

\section{Conclusions}
We have analyzed optical low resolution spectra taken with OSIRIS@GTC of two candidate photoevaporating disks in the area surrounding the massive association Cygnus~OB2. These candidate photoevaporating protostars were recently discovered by \citet{WrightDDG2012} and some have also been found to host T~Tauri stars with disks \citep{GuarcelloDWD2013}. \par

  One of the two targets, the protostar \#5 using the nomenclature by \citet{WrightDDG2012}, hosts a K$2\pm1$ star. Its young age is demonstrated by the presence of a Li~$\lambda$6708 absorption line and evidence of active accretion. Its spectra do not show any emission feature related to intense outflow, but allow us to classify the central object as an actively accreting T~Tauri star. \par

The spectra of the other target, object \#7, do not show absorption features that allow an accurate spectral classification. We adopted a G spectral type based on the lack of Paschen series and TiO bands in the observed spectra. The spectra of this star, taken in two different nights of observations, are characterized by the presence of several emission lines that allow us to study the nature of this peculiar object. The main results are as follows:

\begin{itemize}
 
\item The [OI]~$\lambda$6300/([SII] $\lambda$6716+[SII] $\lambda$6731) ratio is typical of environments with a high level of ionization, such as HII regions irradiated by O stars, and the presence of the OI~$\lambda$8446 line is typical of environments with intense local UV field.
 
\item A strong Ca~II triplet is observed, with a saturated 8498:8542:8662 \AA{}  flux ratio typical of optically thick emission. The mass accretion rate estimated from the luminosity of the Ca~II~$\lambda$8542 line is typical of intense accretors. \par

\item The presence of forbidden emission lines (FELs) demonstrates that the star is characterized by outflow. The asymmetric profile seen in some of the FELs can be explained by a superposition of high-velocity ($\sim400\,$km/s) and low-velocity ($\leq 100\,$km/s) components typical of the presence of both disk wind and stellar jet.

\item The [SII] $\lambda$6716/[SII] $\lambda$6731 ratio is in the saturated regime, corresponding to an electron density in the outflow of $log \left(N_e \right)=4.03\,$cm$^{-3}$.

\item We derived a high mass loss rate from [OI]~$\lambda$6300 and [SII] $\lambda$6731.

\item Both sources show large amplitude and rapid variability both in the continuum and the intensity of the emission lines. In the object \#7 we find evidence suggesting that the variability is due to a rapid increase of infall/outflow activity.

\end{itemize}

We find that in the protostar \#7 the mass loss rate is higher than the mass accretion rate, which is a characteristic of evaporating disks. This result is also supported by the comparison with the well-studied proplyds in Orion, the high level of ionization and the intense local UV field in the circumstellar environment. Even if our data do not allow us to differentiate between the scenarios in which the photoevaporation is externally induced by the OB stars in the center of Cygnus~OB2 or self-induced by the central star itself, with the luminous Lyman continuum produced by the accretion shocks at the stellar surface, our study clearly indicates that the disk in protostar \#7 is photoevaporating as a consequence of intense incident UV radiation. \par


\acknowledgments
We thank the anonymous referee who helped us to significantly improve the quality of this manuscript. This paper is based on observations made with the Gran Telescopio Canarias (GTC), installed in the Spanish Observatorio del Roque de los Muchachos of the Instituto de Astrof\'{i}sica de Canarias, in the island of La Palma. This paper is also based on data from the IPHAS and SDSS Data Release 9. The IPHAS  survey has been carried out at the INT. The INT is operated on the island of La Palma by the Isaac Newton Group in the Spanish Observatorio del Roque de los Muchachos of the Instituto de Astrof\'{i}sica de Canarias. Funding for SDSS-III has been provided by the Alfred P. Sloan Foundation, the Participating Institutions, the National Science Foundation, and the U.S. Department of Energy Office of Science. M. G. G. was supported by the Chandra grant GO0-11040X during the course of this work. J.J.D. was funded by NASA contract NAS8-03060 to the Chandra X-ray Center and thanks the Director, Belinda Wilkes, for continuing support. M. G. G. also acknowledges the grant PRIN-INAF 2012 (P.I. E. Flaccomio). 
%
\addcontentsline{toc}{section}{\bf Bibliografia}
\bibliographystyle{aa}
\bibliography{biblio}

\begin{thebibliography}{101}
\expandafter\ifx\csname natexlab\endcsname\relax\def\natexlab#1{#1}\fi

\bibitem[{{Adams} {et~al.}(2004){Adams}, {Hollenbach}, {Laughlin}, \&
  {Gorti}}]{AdamsHLG2004}
{Adams}, F.~C., {Hollenbach}, D., {Laughlin}, G., \& {Gorti}, U. 2004, \apj,
  611, 360

\bibitem[{{Alexander} \& {Armitage}(2006)}]{AlexanderArmitage2006}
{Alexander}, R.~D. \& {Armitage}, P.~J. 2006, \apjl, 639, L83

\bibitem[{{Alexander} {et~al.}(2006){Alexander}, {Clarke}, \&
  {Pringle}}]{AlexanderCP2006}
{Alexander}, R.~D., {Clarke}, C.~J., \& {Pringle}, J.~E. 2006, \mnras, 369, 216

\bibitem[{{Allen} \& {Strom}(1995)}]{AllenStrom1995}
{Allen}, L.~E. \& {Strom}, K.~M. 1995, \aj, 109, 1379

\bibitem[{{Appenzeller} \& {Wolf}(1977)}]{AppenzellerWolf1977}
{Appenzeller}, I. \& {Wolf}, B. 1977, \aap, 54, 713

\bibitem[{{Bally} {et~al.}(1998){Bally}, {Sutherland}, {Devine}, \&
  {Johnstone}}]{BallySDJ1998}
{Bally}, J., {Sutherland}, R.~S., {Devine}, D., \& {Johnstone}, D. 1998, \aj,
  116, 293

\bibitem[{{Balog} {et~al.}(2007){Balog}, {Muzerolle}, {Rieke}, {Su}, {Young},
  \& {Megeath}}]{BalogMRS2007}
{Balog}, Z., {Muzerolle}, J., {Rieke}, G.~H., {et~al.} 2007, \apj, 660, 1532

\bibitem[{{Balog} {et~al.}(2006){Balog}, {Rieke}, {Su}, {Muzerolle}, \&
  {Young}}]{BalogRSM2006}
{Balog}, Z., {Rieke}, G.~H., {Su}, K.~Y.~L., {Muzerolle}, J., \& {Young}, E.~T.
  2006, \apjl, 650, L83

\bibitem[{{Barentsen} {et~al.}(2011){Barentsen}, {Vink}, {Drew}, {Greimel},
  {Wright}, {Drake}, {Martin}, {Valdivielso}, \& {Corradi}}]{BarentsenVDG2011}
{Barentsen}, G., {Vink}, J.~S., {Drew}, J.~E., {et~al.} 2011, \mnras, 415, 103

\bibitem[{{Beerer} {et~al.}(2010){Beerer}, {Koenig}, {Hora}, {Gutermuth},
  {Bontemps}, {Megeath}, {Schneider}, {Motte}, {Carey}, {Simon}, {Keto},
  {Smith}, {Allen}, {Fazio}, {Kraemer}, {Price}, {Mizuno}, {Adams},
  {Hern{\'a}ndez}, \& {Lucas}}]{BeererKHG2010}
{Beerer}, I.~M., {Koenig}, X.~P., {Hora}, J.~L., {et~al.} 2010, \apj, 720, 679

\bibitem[{{Bowen} \& {Swings}(1947)}]{BowenSwings1947}
{Bowen}, I.~S. \& {Swings}, P. 1947, \apj, 105, 92

\bibitem[{{Brandner} {et~al.}(2000){Brandner}, {Grebel}, {Chu}, {Dottori},
  {Brandl}, {Richling}, {Yorke}, {Points}, \& {Zinnecker}}]{BrandnerGCD2000}
{Brandner}, W., {Grebel}, E.~K., {Chu}, Y.-H., {et~al.} 2000, \aj, 119, 292

\bibitem[{{Cabrit} {et~al.}(1990){Cabrit}, {Edwards}, {Strom}, \&
  {Strom}}]{CabritESS1990}
{Cabrit}, S., {Edwards}, S., {Strom}, S.~E., \& {Strom}, K.~M. 1990, \apj, 354,
  687

\bibitem[{{Calvet} \& {Gullbring}(1998)}]{CalvetGullbring1998}
{Calvet}, N. \& {Gullbring}, E. 1998, \apj, 509, 802

\bibitem[{{Calvet} {et~al.}(2000){Calvet}, {Hartmann}, \&
  {Strom}}]{CalvetHartmannStroom2000}
{Calvet}, N., {Hartmann}, L., \& {Strom}, S.~E. 2000, Protostars and Planets
  IV, 377

\bibitem[{{Cardelli} {et~al.}(1989){Cardelli}, {Clayton}, \&
  {Mathis}}]{CardelliCM1989}
{Cardelli}, J.~A., {Clayton}, G.~C., \& {Mathis}, J.~S. 1989, \apj, 345, 245

\bibitem[{{Cepa} {et~al.}(2000){Cepa}, {Aguiar}, \& {Escalera}}]{CepaAEG2000}
{Cepa}, J., {Aguiar}, M., \& {Escalera}, e.~a. 2000, in Presented at the
  Society of Photo-Optical Instrumentation Engineers (SPIE) Conference, Vol.
  4008, Society of Photo-Optical Instrumentation Engineers (SPIE) Conference
  Series, ed. {M.~Iye \& A.~F.~Moorwood}, 623--631

\bibitem[{{Chou} {et~al.}(2013){Chou}, {Takami}, {Manset}, {Beck}, {Pyo},
  {Chen}, {Panwar}, {Karr}, {Shang}, \& {Liu}}]{ChouTMB2013}
{Chou}, M.-Y., {Takami}, M., {Manset}, N., {et~al.} 2013, \aj, 145, 108

\bibitem[{{Cody} {et~al.}(2014){Cody}, {Stauffer}, {Baglin}, {Micela},
  {Rebull}, {Flaccomio}, {Morales-Calder{\'o}n}, {Aigrain}, {Bouvier},
  {Hillenbrand}, {Gutermuth}, {Song}, {Turner}, {Alencar}, {Zwintz},
  {Plavchan}, {Carpenter}, {Findeisen}, {Carey}, {Terebey}, {Hartmann},
  {Calvet}, {Teixeira}, {Vrba}, {Wolk}, {Covey}, {Poppenhaeger}, {G{\"u}nther},
  {Forbrich}, {Whitney}, {Affer}, {Herbst}, {Hora}, {Barrado}, {Holtzman},
  {Marchis}, {Wood}, {Medeiros Guimar{\~a}es}, {Lillo Box}, {Gillen},
  {McQuillan}, {Espaillat}, {Allen}, {D'Alessio}, \& {Favata}}]{CodySBM2014}
{Cody}, A.~M., {Stauffer}, J., {Baglin}, A., {et~al.} 2014, \aj, 147, 82

\bibitem[{{Comer{\'o}n} {et~al.}(2002){Comer{\'o}n}, {Pasquali}, {Rodighiero},
  {Stanishev}, {De Filippis}, {L{\'o}pez Mart{\'{\i}}}, {G{\'a}lvez Ortiz},
  {Stankov}, \& {Gredel}}]{ComeronPRS2002}
{Comer{\'o}n}, F., {Pasquali}, A., {Rodighiero}, G., {et~al.} 2002, \aap, 389,
  874

\bibitem[{{Contreras Pe{\~n}a} {et~al.}(2014){Contreras Pe{\~n}a}, {Lucas},
  {Froebrich}, {Kumar}, {Goldstein}, {Drew}, {Adamson}, {Davis}, {Barentsen},
  \& {Wright}}]{ContrerasPenaLFK2014}
{Contreras Pe{\~n}a}, C., {Lucas}, P.~W., {Froebrich}, D., {et~al.} 2014,
  \mnras, 439, 1829

\bibitem[{{Czyzak} {et~al.}(1986){Czyzak}, {Keyes}, \& {Aller}}]{CzyzakKA1986}
{Czyzak}, S.~J., {Keyes}, C.~D., \& {Aller}, L.~H. 1986, \apjs, 61, 159

\bibitem[{{Drake}(2014, in preparation)}]{Drakeinprep}
{Drake}, J.~J. 2014, in preparation, \apj

\bibitem[{{Drew} {et~al.}(2005){Drew}, {Greimel}, {Irwin}, {Aungwerojwit},
  {Barlow}, {Corradi}, {Drake}, {G{\"a}nsicke}, {Groot}, {Hales}, {Hopewell},
  {Irwin}, {Knigge}, {Leisy}, {Lennon}, {Mampaso}, {Masheder}, {Matsuura},
  {Morales-Rueda}, {Morris}, {Parker}, {Phillipps}, {Rodriguez-Gil}, {Roelofs},
  {Skillen}, {Sokoloski}, {Steeghs}, {Unruh}, {Viironen}, {Vink}, {Walton},
  {Witham}, {Wright}, {Zijlstra}, \& {Zurita}}]{DrewGIA2005}
{Drew}, J.~E., {Greimel}, R., {Irwin}, M.~J., {et~al.} 2005, \mnras, 362, 753

\bibitem[{{Edwards} {et~al.}(1994){Edwards}, {Hartigan}, {Ghandour}, \&
  {Andrulis}}]{EdwardsHGA1994}
{Edwards}, S., {Hartigan}, P., {Ghandour}, L., \& {Andrulis}, C. 1994, \aj,
  108, 1056

\bibitem[{{Elmegreen}(2011)}]{Elmegreen2011}
{Elmegreen}, B.~G. 2011, in EAS Publications Series, Vol.~51, EAS Publications
  Series, ed. C.~{Charbonnel} \& T.~{Montmerle}, 45--58

\bibitem[{{Fang} {et~al.}(2012){Fang}, {van Boekel}, {King}, {Henning},
  {Bouwman}, {Doi}, {Okamoto}, {Roccatagliata}, \&
  {Sicilia-Aguilar}}]{FangBKH2012}
{Fang}, M., {van Boekel}, R., {King}, R.~R., {et~al.} 2012, \aap, 539, A119

\bibitem[{{Fedele} {et~al.}(2010){Fedele}, {van den Ancker}, {Henning},
  {Jayawardhana}, \& {Oliveira}}]{FedeleVHJ2010}
{Fedele}, D., {van den Ancker}, M.~E., {Henning}, T., {Jayawardhana}, R., \&
  {Oliveira}, J.~M. 2010, \aap, 510, A72

\bibitem[{{Grandi}(1975)}]{Grandi1975}
{Grandi}, S.~A. 1975, \apj, 196, 465

\bibitem[{{Grandi}(1980)}]{Grandi1980}
{Grandi}, S.~A. 1980, \apj, 238, 10

\bibitem[{{Griffith} {et~al.}(1991){Griffith}, {Heflin}, {Conner}, {Burke}, \&
  {Langston}}]{GriffithHCB1991}
{Griffith}, M., {Heflin}, M., {Conner}, S., {Burke}, B., \& {Langston}, G.
  1991, \apjs, 75, 801

\bibitem[{{Guarcello}(in preparation)}]{Guarcelloinprep1}
{Guarcello}, M.~G. in preparation, \apj

\bibitem[{{Guarcello} {et~al.}(2013){Guarcello}, {Drake}, {Wright}, {Drew},
  {Gutermuth}, {Hora}, {Naylor}, {Aldcroft}, {Fruscione},
  {Garc{\'{\i}}a-Alvarez}, {Kashyap}, \& {King}}]{GuarcelloDWD2013}
{Guarcello}, M.~G., {Drake}, J.~J., {Wright}, N.~J., {et~al.} 2013, \apj, 773,
  135

\bibitem[{{Guarcello} {et~al.}(2009){Guarcello}, {Micela}, {Damiani}, {Peres},
  {Prisinzano}, \& {Sciortino}}]{GuarcelloMDP2009}
{Guarcello}, M.~G., {Micela}, G., {Damiani}, F., {et~al.} 2009, \aap, 496, 453

\bibitem[{{Guarcello} {et~al.}(2010){Guarcello}, {Micela}, {Peres},
  {Prisinzano}, \& {Sciortino}}]{GuarcelloMPP2010}
{Guarcello}, M.~G., {Micela}, G., {Peres}, G., {Prisinzano}, L., \&
  {Sciortino}, S. 2010, \aap, 521, A61

\bibitem[{{Guarcello} {et~al.}(2007){Guarcello}, {Prisinzano}, {Micela},
  {Damiani}, {Peres}, \& {Sciortino}}]{GuarcelloPMD2007}
{Guarcello}, M.~G., {Prisinzano}, L., {Micela}, G., {et~al.} 2007, \aap, 462,
  245

\bibitem[{{Guarcello} {et~al.}(2012){Guarcello}, {Wright}, {Drake},
  {Garc{\'{\i}}a-Alvarez}, {Drew}, {Aldcroft}, \& {Kashyap}}]{GuarcelloWDG2012}
{Guarcello}, M.~G., {Wright}, N.~J., {Drake}, J.~J., {et~al.} 2012, \apjs, 202,
  19

\bibitem[{{Gullbring} {et~al.}(1998){Gullbring}, {Hartmann}, {Briceno}, \&
  {Calvet}}]{GullbringHBC1998}
{Gullbring}, E., {Hartmann}, L., {Briceno}, C., \& {Calvet}, N. 1998, \apj,
  492, 323

\bibitem[{{Gutermuth} {et~al.}(2008){Gutermuth}, {Myers}, {Megeath}, {Allen},
  {Pipher}, {Muzerolle}, {Porras}, {Winston}, \& {Fazio}}]{GutermuthMMA2008}
{Gutermuth}, R.~A., {Myers}, P.~C., {Megeath}, S.~T., {et~al.} 2008, \apj, 674,
  336

\bibitem[{{Habing}(1968)}]{Habing1968}
{Habing}, H.~J. 1968, \bain, 19, 421

\bibitem[{{Haisch} {et~al.}(2001){Haisch}, {Lada}, \& {Lada}}]{HaischLL2001}
{Haisch}, Jr., K.~E., {Lada}, E.~A., \& {Lada}, C.~J. 2001, \apjl, 553, L153

\bibitem[{{Hamann}(1994)}]{Hamann1994}
{Hamann}, F. 1994, \apjs, 93, 485

\bibitem[{{Hamann} \& {Persson}(1992)}]{HamannPersson1992}
{Hamann}, F. \& {Persson}, S.~E. 1992, \apjs, 82, 247

\bibitem[{{Hartigan} {et~al.}(1995){Hartigan}, {Edwards}, \&
  {Ghandour}}]{HartiganEG1995}
{Hartigan}, P., {Edwards}, S., \& {Ghandour}, L. 1995, \apj, 452, 736

\bibitem[{{Hartmann} {et~al.}(1998){Hartmann}, {Calvet}, {Gullbring}, \&
  {D'Alessio}}]{HartmannCGD1998}
{Hartmann}, L., {Calvet}, N., {Gullbring}, E., \& {D'Alessio}, P. 1998, \apj,
  495, 385

\bibitem[{{Henney} \& {Arthur}(1998)}]{HenneyArthur1998}
{Henney}, W.~J. \& {Arthur}, S.~J. 1998, \aj, 116, 322

\bibitem[{{Henney} \& {O'Dell}(1999)}]{HenneyODell1999}
{Henney}, W.~J. \& {O'Dell}, C.~R. 1999, \aj, 118, 2350

\bibitem[{{Herbig}(1966)}]{Herbig1966}
{Herbig}, G.~H. 1966, Vistas in Astronomy, 8, 109

\bibitem[{{Herbig} \& {Soderblom}(1980)}]{HerbigSoderblom1980}
{Herbig}, G.~H. \& {Soderblom}, D.~R. 1980, \apj, 242, 628

\bibitem[{{Hern{\'a}ndez} {et~al.}(2004){Hern{\'a}ndez}, {Calvet},
  {Brice{\~n}o}, {Hartmann}, \& {Berlind}}]{HernandezCBH2004}
{Hern{\'a}ndez}, J., {Calvet}, N., {Brice{\~n}o}, C., {Hartmann}, L., \&
  {Berlind}, P. 2004, \aj, 127, 1682

\bibitem[{{Hern{\'a}ndez} {et~al.}(2007){Hern{\'a}ndez}, {Hartmann}, {Megeath},
  {Gutermuth}, {Muzerolle}, {Calvet}, {Vivas}, {Brice{\~n}o}, {Allen},
  {Stauffer}, {Young}, \& {Fazio}}]{HernandezHMG2007}
{Hern{\'a}ndez}, J., {Hartmann}, L., {Megeath}, T., {et~al.} 2007, \apj, 662,
  1067

\bibitem[{{Hester} {et~al.}(1996){Hester}, {Scowen}, {Sankrit}, {Lauer},
  {Ajhar}, {Baum}, {Code}, {Currie}, {Danielson}, {Ewald}, {Faber},
  {Grillmair}, {Groth}, {Holtzman}, {Hunter}, {Kristian}, {Light}, {Lynds},
  {Monet}, {O'Neil}, {Shaya}, {Seidelmann}, \& {Westphal}}]{HesterSSL1996}
{Hester}, J.~J., {Scowen}, P.~A., {Sankrit}, R., {et~al.} 1996, \aj, 111, 2349

\bibitem[{{Hirth}(1994)}]{Hirth1994}
{Hirth}, G.~A. 1994, in Astronomical Society of the Pacific Conference Series,
  Vol.~57, Stellar and Circumstellar Astrophysics, a 70th birthday celebration
  for K. H. Bohm and E. Bohm-Vitense, ed. G.~{Wallerstein} \&
  A.~{Noriega-Crespo}, 32

\bibitem[{{Hirth} {et~al.}(1997){Hirth}, {Mundt}, \& {Solf}}]{HirthMS1997}
{Hirth}, G.~A., {Mundt}, R., \& {Solf}, J. 1997, \aaps, 126, 437

\bibitem[{{Jeffries} \& {Oliveira}(2005)}]{JeffriesOliveira2005}
{Jeffries}, R.~D. \& {Oliveira}, J.~M. 2005, \mnras, 358, 13

\bibitem[{{Johansson} \& {Letokhov}(2005)}]{JohanssonLetokhov2005}
{Johansson}, S. \& {Letokhov}, V.~S. 2005, \mnras, 364, 731

\bibitem[{{Johnstone} {et~al.}(1998){Johnstone}, {Hollenbach}, \&
  {Bally}}]{JohnstoneHB1998}
{Johnstone}, D., {Hollenbach}, D., \& {Bally}, J. 1998, \apj, 499, 758

\bibitem[{{Kastner} \& {Bhatia}(1995)}]{KastnerBhatia1995}
{Kastner}, S.~O. \& {Bhatia}, A.~K. 1995, \apj, 439, 346

\bibitem[{{Kepner} {et~al.}(1993){Kepner}, {Hartigan}, {Yang}, \&
  {Strom}}]{KepnerHYS1993}
{Kepner}, J., {Hartigan}, P., {Yang}, C., \& {Strom}, S. 1993, \apjl, 415, L119

\bibitem[{{Koenig} {et~al.}(2008){Koenig}, {Allen}, {Kenyon}, {Su}, \&
  {Balog}}]{KoenigAKS2008}
{Koenig}, X.~P., {Allen}, L.~E., {Kenyon}, S.~J., {Su}, K.~Y.~L., \& {Balog},
  Z. 2008, \apjl, 687, L37

\bibitem[{{Kwan}(1997)}]{Kwan1997}
{Kwan}, J. 1997, \apj, 489, 284

\bibitem[{{Kwan} \& {Fischer}(2011)}]{KwanFischer2011}
{Kwan}, J. \& {Fischer}, W. 2011, \mnras, 411, 2383

\bibitem[{{Kwan} \& {Tademaru}(1988)}]{KwanTademaru1988}
{Kwan}, J. \& {Tademaru}, E. 1988, \apjl, 332, L41

\bibitem[{{Lada} {et~al.}(2006){Lada}, {Muench}, {Luhman}, {Allen}, {Hartmann},
  {Megeath}, {Myers}, {Fazio}, {Wood}, {Muzerolle}, {Rieke}, {Siegler}, \&
  {Young}}]{LadaMLA2006}
{Lada}, C.~J., {Muench}, A.~A., {Luhman}, K.~L., {et~al.} 2006, \aj, 131, 1574

\bibitem[{{Mamajek}(2009)}]{Mamajek2009}
{Mamajek}, E.~E. 2009, in American Institute of Physics Conference Series, Vol.
  1158, American Institute of Physics Conference Series, ed. T.~{Usuda},
  M.~{Tamura}, \& M.~{Ishii}, 3--10

\bibitem[{{Martins} {et~al.}(2005){Martins}, {Schaerer}, \&
  {Hillier}}]{MartinSH2005}
{Martins}, F., {Schaerer}, D., \& {Hillier}, D.~J. 2005, \aap, 436, 1049

\bibitem[{{Mathew} {et~al.}(2012){Mathew}, {Banerjee}, {Subramaniam}, \&
  {Ashok}}]{MathewBSA2012}
{Mathew}, B., {Banerjee}, D.~P.~K., {Subramaniam}, A., \& {Ashok}, N.~M. 2012,
  \apj, 753, 13

\bibitem[{{McCaughrean} \& {Andersen}(2002)}]{McCaughreanAndersen2002}
{McCaughrean}, M.~J. \& {Andersen}, M. 2002, \aap, 389, 513

\bibitem[{{Morales-Calder{\'o}n} {et~al.}(2011){Morales-Calder{\'o}n},
  {Stauffer}, {Hillenbrand}, {Gutermuth}, {Song}, {Rebull}, {Plavchan},
  {Carpenter}, {Whitney}, {Covey}, {Alves de Oliveira}, {Winston},
  {McCaughrean}, {Bouvier}, {Guieu}, {Vrba}, {Holtzman}, {Marchis}, {Hora},
  {Wasserman}, {Terebey}, {Megeath}, {Guinan}, {Forbrich}, {Hu{\'e}lamo},
  {Riviere-Marichalar}, {Barrado}, {Stapelfeldt}, {Hern{\'a}ndez}, {Allen},
  {Ardila}, {Bayo}, {Favata}, {James}, {Werner}, \&
  {Wood}}]{Morales-CalderonSHG2011}
{Morales-Calder{\'o}n}, M., {Stauffer}, J.~R., {Hillenbrand}, L.~A., {et~al.}
  2011, \apj, 733, 50

\bibitem[{{M{\"u}nch} \& {Taylor}(1974)}]{MuenchTaylor1974}
{M{\"u}nch}, G. \& {Taylor}, K. 1974, \apjl, 192, L93

\bibitem[{{Muzerolle} {et~al.}(1998){Muzerolle}, {Hartmann}, \&
  {Calvet}}]{MuzerolleHC1998}
{Muzerolle}, J., {Hartmann}, L., \& {Calvet}, N. 1998, \aj, 116, 455

\bibitem[{{Natta} {et~al.}(2004){Natta}, {Testi}, {Muzerolle}, {Randich},
  {Comer{\'o}n}, \& {Persi}}]{NattaTMR2004}
{Natta}, A., {Testi}, L., {Muzerolle}, J., {et~al.} 2004, \aap, 424, 603

\bibitem[{{Parravano} {et~al.}(2003){Parravano}, {Hollenbach}, \&
  {McKee}}]{ParravanoHM2003}
{Parravano}, A., {Hollenbach}, D.~J., \& {McKee}, C.~F. 2003, \apj, 584, 797

\bibitem[{{Pereira} \& {Miranda}(2007)}]{PereiraMiranda2007}
{Pereira}, C.~B. \& {Miranda}, L.~F. 2007, \aap, 462, 231

\bibitem[{{Proxauf} {et~al.}(2014){Proxauf}, {{\"O}ttl}, \&
  {Kimeswenger}}]{ProxaufOK2014}
{Proxauf}, B., {{\"O}ttl}, S., \& {Kimeswenger}, S. 2014, \aap, 561, A10

\bibitem[{{Randich} {et~al.}(2005){Randich}, {Bragaglia}, {Pastori},
  {Prisinzano}, {Sestito}, {Spano}, {Villanova}, {Carraro}, {Carretta},
  {Romano}, {Zaggia}, {Pallavicini}, {Pasquini}, {Primas}, {Tagliaferri}, \&
  {Tosi}}]{RandichBPP2005}
{Randich}, S., {Bragaglia}, A., {Pastori}, L., {et~al.} 2005, The Messenger,
  121, 18

\bibitem[{{Rigliaco} {et~al.}(2009){Rigliaco}, {Natta}, {Randich}, \&
  {Sacco}}]{RigliacoNRS2009}
{Rigliaco}, E., {Natta}, A., {Randich}, S., \& {Sacco}, G. 2009, \aap, 495, L13

\bibitem[{{Robitaille} {et~al.}(2007){Robitaille}, {Whitney}, {Indebetouw}, \&
  {Wood}}]{RobitailleWI2007}
{Robitaille}, T.~P., {Whitney}, B.~A., {Indebetouw}, R., \& {Wood}, K. 2007,
  \apjs, 169, 328

\bibitem[{{Rygl} {et~al.}(2012){Rygl}, {Brunthaler}, {Sanna}, {Menten}, {Reid},
  {van Langevelde}, {Honma}, {Torstensson}, \& {Fujisawa}}]{RyglBSM2012}
{Rygl}, K.~L.~J., {Brunthaler}, A., {Sanna}, A., {et~al.} 2012, \aap, 539, A79

\bibitem[{{Siess} {et~al.}(2000){Siess}, {Dufour}, \&
  {Forestini}}]{SiessDF2000}
{Siess}, L., {Dufour}, E., \& {Forestini}, M. 2000, \aap, 358, 593

\bibitem[{{Smith} {et~al.}(2003){Smith}, {Bally}, \& {Morse}}]{SmithBM2003}
{Smith}, N., {Bally}, J., \& {Morse}, J.~A. 2003, \apjl, 587, L105

\bibitem[{{Solf}(1989)}]{Solf1989}
{Solf}, J. 1989, in European Southern Observatory Conference and Workshop
  Proceedings, Vol.~33, European Southern Observatory Conference and Workshop
  Proceedings, ed. B.~{Reipurth}, 399--406

\bibitem[{{Stauffer} {et~al.}(2014){Stauffer}, {Cody}, {Baglin}, {Alencar},
  {Rebull}, {Hillenbrand}, {Venuti}, {Turner}, {Carpenter}, {Plavchan},
  {Findeisen}, {Carey}, {Terebey}, {Morales-Calder{\'o}n}, {Bouvier}, {Micela},
  {Flaccomio}, {Song}, {Gutermuth}, {Hartmann}, {Calvet}, {Whitney}, {Barrado},
  {Vrba}, {Covey}, {Herbst}, {Furesz}, {Aigrain}, \&
  {Favata}}]{StaufferCBA2014}
{Stauffer}, J., {Cody}, A.~M., {Baglin}, A., {et~al.} 2014, \aj, 147, 83

\bibitem[{{Stecklum} {et~al.}(1998){Stecklum}, {Henning}, {Feldt}, {Hayward},
  {Hoare}, {Hofner}, \& {Richter}}]{StecklumHFH1998}
{Stecklum}, B., {Henning}, T., {Feldt}, M., {et~al.} 1998, \aj, 115, 767

\bibitem[{{Storzer} \& {Hollenbach}(1998)}]{StorzerHollenbach1998}
{Storzer}, H. \& {Hollenbach}, D. 1998, \apjl, 502, L71

\bibitem[{{St{\"o}rzer} \& {Hollenbach}(1999)}]{StorzerHollenbach1999}
{St{\"o}rzer}, H. \& {Hollenbach}, D. 1999, \apj, 515, 669

\bibitem[{{St{\"o}rzer} \& {Hollenbach}(2000)}]{StorzerHollenbach2000}
{St{\"o}rzer}, H. \& {Hollenbach}, D. 2000, \apj, 539, 751

\bibitem[{{Strelnitski} {et~al.}(2013){Strelnitski}, {Bieging}, {Hora},
  {Smith}, {Armstrong}, {Lagergren}, \& {Walker}}]{StrelnitskiBHS2013}
{Strelnitski}, V., {Bieging}, J.~H., {Hora}, J., {et~al.} 2013, \apj, 777, 89

\bibitem[{{Throop} \& {Bally}(2005)}]{ThroopBally2005}
{Throop}, H.~B. \& {Bally}, J. 2005, \apjl, 623, L149

\bibitem[{{Torres-Dodgen} \& {Weaver}(1993)}]{Torres-DodgenWeaver1993}
{Torres-Dodgen}, A.~V. \& {Weaver}, W.~B. 1993, \pasp, 105, 693

\bibitem[{{Uchida} \& {Shibata}(1984)}]{UchidaShibata1984}
{Uchida}, Y. \& {Shibata}, K. 1984, \pasj, 36, 105

\bibitem[{{Vink} {et~al.}(2008){Vink}, {Drew}, {Steeghs}, {Wright}, {Martin},
  {G{\"a}nsicke}, {Greimel}, \& {Drake}}]{Vink2008}
{Vink}, J.~S., {Drew}, J.~E., {Steeghs}, D., {et~al.} 2008, \mnras, 387, 308

\bibitem[{{Walborn} \& {Fitzpatrick}(1990)}]{WalbornFitzpatrick1990}
{Walborn}, N.~R. \& {Fitzpatrick}, E.~L. 1990, \pasp, 102, 379

\bibitem[{{Whelan} {et~al.}(2004){Whelan}, {Ray}, \& {Davis}}]{WhelanRD2004}
{Whelan}, E.~T., {Ray}, T.~P., \& {Davis}, C.~J. 2004, \aap, 417, 247

\bibitem[{{Whitworth}(1979)}]{Whitworth1979}
{Whitworth}, A. 1979, \mnras, 186, 59

\bibitem[{{Wiese} {et~al.}(1969){Wiese}, {Smith}, \&
  {Miles}}]{WieseSmithMiles1969}
{Wiese}, W.~L., {Smith}, M.~W., \& {Miles}, B.~M. 1969, {Atomic transition
  probabilities. Vol. 2: Sodium through Calcium. A critical data compilation}

\bibitem[{{Wright}(2014, in preparation)}]{Wrightinprep}
{Wright}, N.~J. 2014, in preparation, \apj

\bibitem[{{Wright} {et~al.}(2012){Wright}, {Drake}, {Drew}, {Guarcello},
  {Gutermuth}, {Hora}, \& {Kraemer}}]{WrightDDG2012}
{Wright}, N.~J., {Drake}, J.~J., {Drew}, J.~E., {et~al.} 2012, \apjl, 746, L21

\bibitem[{{Wright} {et~al.}(2010){Wright}, {Drake}, {Drew}, \&
  {Vink}}]{WrightDDV2010}
{Wright}, N.~J., {Drake}, J.~J., {Drew}, J.~E., \& {Vink}, J.~S. 2010, \apj,
  713, 871

\bibitem[{{Wright} {et~al.}(2014){Wright}, {Parker}, {Goodwin}, \&
  {Drake}}]{WrightPGD2014}
{Wright}, N.~J., {Parker}, R.~J., {Goodwin}, S.~P., \& {Drake}, J.~J. 2014,
  \mnras, 438, 639

\bibitem[{{Yusef-Zadeh} {et~al.}(2005){Yusef-Zadeh}, {Biretta}, \&
  {Geballe}}]{Yusef-ZadehBG2005}
{Yusef-Zadeh}, F., {Biretta}, J., \& {Geballe}, T.~R. 2005, \aj, 130, 1171

\end{thebibliography}


        \begin{table*}[]
        \centering
        \caption {Photometry of the central star in the protostar \#5}
        \vspace{0.5cm}

        \begin{tabular}{ccccc}
        \hline
        \hline
        \multicolumn{5}{c}{SDSS} \\
        \hline
        u (mag) & g (mag) & r (mag) & i (mag) & z (mag)  \\
        \hline
	$24\pm 1$ & $23.7\pm0.9$ & $19.8\pm 0.4$ & $18.0\pm0.3$ & $16.4\pm0.2$ \\
	\hline	
        \hline
	\end{tabular}

        \begin{tabular}{ccc}
        \hline
        \hline
        \multicolumn{3}{c}{IPHAS} \\
        \hline
        r (mag) & i (mag) & H$\alpha$ (mag)  \\
        \hline
	$19.7\pm 0.02$ & $17.47\pm0.02$ & $18.07\pm0.02$ \\
	\hline	
        \hline
	\end{tabular}

        \begin{tabular}{ccc}
        \hline
        \hline
        \multicolumn{3}{c}{UKIDSS} \\
        \hline
        J (mag) & H (mag) & K (mag)  \\
        \hline
	$13.395\pm 0.001$ & $11.894\pm0.001$ & $15.513\pm 0.001$ \\
	\hline	
        \hline
	\end{tabular}

	\begin{tabular}{cccc}
        \hline
        \hline
        \multicolumn{4}{c}{Spitzer} \\
        \hline
        [3.6] (mag) & [4.5] (mag) & [5.8] (mag) & [8.0] (mag)  \\
        \hline
	$8.648\pm 0.007$&	$7.824\pm0.004$&	$7.078\pm0.002$&	$6.16\pm0.04$\\
	\hline	
        \hline
        \end{tabular}
        \label{prop5_tb}
        \end{table*}

        \begin{table*}[]
        \centering
        \caption {Log of the observations}
        \vspace{0.5cm}
        \begin{tabular}{ccccccc}
        \hline
        \hline
        OB &  Target &  Date& Grism  &  Seeing  & Exposure Time & Standard  \\
        \hline
	OB1&	protostar \#5&	2012-08-10&	R1000B&			$\leq1.25^{\prime\prime}$&	$3\times 180\,s$&	L1363-3\\
	OB2&	protostar \#7&	2012-08-10&	R1000R&			$\leq1.16^{\prime\prime}$&	$2\times 870\,s$&	L1363-3\\
	OB3&	protostar \#5&	2012-08-14&	R2500R, R2500I&		$\leq1.09^{\prime\prime}$&	$2\times 870\,s$&	G158-100\\
	OB4&	protostar \#7&	2012-08-14&	R1000R&			$\leq1.31^{\prime\prime}$&	$4\times 870\,s$&	G158-100\\
	\hline
        \hline
        \multicolumn{7}{l}{} 
        \end{tabular}
        \label{log_tb}
        \end{table*}

       \begin{table}[]
        \centering
        \caption {Width and central wavelength of the used grisms}
        \vspace{0.5cm}
        \begin{tabular}{ccc}
        \hline
        \hline
        Grism &  $\lambda_c$ (\AA{}) &  spectral range (\AA{})  \\
        \hline
	R1000R&	7510&	5100-10000\\
	R1000B&	5510&	3630-7500 \\
	R2500R&	6590&	5575-7685 \\
	R2500I&	8740&	7330-10000\\
	\hline
        \hline
        \multicolumn{3}{l}{} 
        \end{tabular}
        \label{grism_tb}
        \end{table}

       \begin{table}[]
        \centering
        \caption {Centroid velocity of the observed FELs}
        \vspace{0.5cm}
        \begin{tabular}{ccc}
        \hline
        \hline
        Line &  OB2 velocity  &  OB4 velocity\\
	     & $\left(km/s \right)$ & $\left(km/s \right)$ \\
        \hline
	$[$OI$]$5577&		$-174\pm12$&	$-62\pm3$\\
	$[$OI$]$6300&		$-47\pm8$&	$-210\pm9$\\
	$[$OI$]$6363&		NAN&		$-65\pm15$\\
	$[$SII$]$6716&		NAN&		$-178\pm20$\\
	$[$SII$]$6731&		$-307\pm16$&	$-249\pm14$\\
	$[$Fe~II$]$7155&	$-102\pm41$&	$-130\pm73$\\
	$[$Fe~II$]$8617&	NAN&		$-31\pm7$\\
	\hline
        \hline
        \multicolumn{3}{l}{} 
        \end{tabular}
        \label{shift_tb}
        \end{table}

       \begin{table*}[]
        \centering
        \caption {Fluxes of the emission lines observed in the two OBs in the protostar \#7 (unreddened spectra).}
        \vspace{0.5cm}
        \begin{tabular}{cccc}
        \hline
        \hline
        Line &  OB2 flux &  OB4 flux& OB4/OB2 ratio  \\
	     &  $\times10^{-15}\,$erg$\,$cm$^{-2}\,$s$^{-1}$  &  $\times10^{-15}\,$erg$\,$cm$^{-2}\,$s$^{-1}$&   \\
        \hline
	$[$OI$]$5577	&	0.05	&	3.7	&	67.9\\
	$[$OI$]$6300	&	0.19	&	1.81	&	9.7\\
	$[$OI$]$6363	&	0.04	&	0.57	&	15.1\\
	$[$SII$]$6716	&	NAN	&	0.97	&	NAN\\
	$[$SII$]$6731	&	0.09	&	2.03	&	23.4\\
	$[$Fe~II$]$7155	&	0.17	&	0.77	&	4.5\\
	$[$Ca~I$]$8498	&	1.38	&	5.51	&	4.0\\
	$[$Ca~I$]$8452	&	1.39	&	5.83	&	4.2\\
	$[$Ca~I$]$8662	&	1.46	&	5.41	&	3.9\\ 
	$[$OI$]$8446	&	0.35	&	1.34	&	3.8\\
	\hline
        \hline
        \multicolumn{4}{l}{} 
        \end{tabular}
        \label{flux_tb}
        \end{table*}

	\begin{table}[]
        \centering
        \caption {Effect of the extinction on the used diagnostics (OB4)}
        \vspace{0.5cm}
        \begin{tabular}{cccc}
        \hline
        \hline
        A$_V$ &  mass loss rate &  mass accretion rate & [OI]/[SII]  \\
	  $m$ &  M$_{\odot}$/yr &  M$_{\odot}$/yr      &                   \\
        \hline
        6	&$5.7\times10^{-7}$	&$1.7\times10^{-8}$	&0.63	\\
        8	&$2.1\times10^{-6}$	&$4.8\times10^{-7}$	&0.94	\\
        12	&$5.0\times10^{-5}$	&$3.9\times10^{-6}$	&1.25	\\
	\hline
        \hline
        \multicolumn{4}{l}{} 
        \end{tabular}
        \label{ext_tb}
        \end{table}

%

\newpage

\end{document}